\documentclass[a4paper,useAMS,usenatbib]{mnras}
\usepackage{graphics,epsfig,psfig}
\usepackage{todonotes}
\usepackage[normalem]{ulem}
\usepackage{xcolor}
\usepackage{float}
\usepackage{subfig}
\usepackage{graphicx}
\usepackage[]{inputenc,amssymb}
\usepackage{natbib}
\usepackage{url}
\usepackage{widetext}
\usepackage{amsmath}
\usepackage{cancel}

\usepackage[style=base, singlelinecheck=off, font=small, labelfont=bf, justification=raggedright]{caption}

  \makeatletter
\def\@fnsymbol#1{\ensuremath{\ifcase#1\or \dagger\or \ddagger\or
   \mathsection\or \mathparagraph\or \|\or **\or \dagger\dagger
   \or \ddagger\ddagger \else\@ctrerr\fi}}
    \makeatother
\def \be{\begin{equation}}
\def \ee{\end{equation}}
\def \bea{\begin{eqnarray}}
\def \eea{\end{eqnarray}}

\def\mnras{MNRAS}
\def\prd{PRD}
\def\aap{A\&A}
\definecolor{webgreen}{rgb}{0,.5,0}
\definecolor{webbrown}{rgb}{.6,0,0}

\def\aap{A\&A}

\def\apjs{ApJS}
\def\apjl{ApJL}
\def\mnras{MNRAS}

\def\prd{Phys.Rev.D}         
\def\araa{ARA\&A}       

\title[Inferring the lensing rates using SGWB]{Inferring the lensing rate of LIGO-Virgo sources from the stochastic gravitational wave background}
\author[Mukherjee, Broadhurst, Diego, Silk, \& Smoot (2020)]
{Suvodip Mukherjee$^{1,2,3}$\thanks{s.mukherjee@uva.nl, mukherje@iap.fr}, Tom  Broadhurst$^{4,5,6}$\thanks{tom.j.broadhurst@gmail.com}, Jose M. Diego$^{7}$\thanks{jdiego@ifca.unican.es}, Joseph Silk$^{2,3,8,9}$\thanks{silk@iap.fr}, 
\newauthor
\& George F. Smoot$^{10, 11, 12}$\thanks{gfsmoot@lbl.gov}\thanks{Author list is in the alphabetical order except the corresponding author.}\\
$^{1}$ Gravitation Astroparticle Physics Amsterdam (GRAPPA),
Anton Pannekoek Institute for Astronomy and Institute for High-Energy Physics,\\
University of Amsterdam, Science Park 904, 1090 GL Amsterdam, The Netherlands\\
$^{2}$Institut d'Astrophysique de Paris (IAP), UMR 7095, CNRS/UPMC Universit\'e Paris 6, Sorbonne Universit\'es,\\ 98 bis boulevard Arago, F-75014 Paris, France\\
$^{3}$ Institut Lagrange de Paris (ILP), Sorbonne Universit\'es, 98 bis Boulevard Arago, 75014 Paris, France\\
$^{4}$ Department of Theoretical Physics, University of the Basque Country UPV-EHU, 48040 Bilbao, Spain\\
$^5$ Donostia International Physics Center (DIPC), 20018 Donostia, The Basque Country\\
$^6$ IKERBASQUE, Basque Foundation for Science, Alameda Urquijo, $36-5$ 48008 Bilbao, Spain\\
$^7$ Instituto de F\'isica de Cantabria, CSIC-Universidad de Cantabria, E-39005 Santander, Spain\\
$^8$ The Johns Hopkins University, Department of Physics \& Astronomy, 3400 N. Charles Street, Baltimore, MD 21218, USA\\
$^{9}$ Beecroft Institute for Cosmology and Particle Astrophysics, University of Oxford, Keble Road, Oxford OX1 3RH, UK\\
$^{10}$ IAS TT \& WF Chao Foundation Professor, IAS, Hong Kong University of Science and Technology,\\
Clear Water Bay, Kowloon, 999077 Hong Kong, China\\
$^{11}$ Paris Centre for Cosmological Physics, Universit\'e de Paris,
CNRS, Astroparticule et Cosmologie, $F-75013$ Paris, France A,\\
10 rue Alice Domon et Leonie Duquet, 75205 Paris CEDEX 13, France\\
$^{12}$  Physics Department and Lawrence Berkeley National Laboratory, University of California, Berkeley, 94720 CA, USA\\
}
\pubyear{2020}
\begin{document}
\label{firstpage}
\pagerange{\pageref{firstpage}--\pageref{lastpage}}
\maketitle
\begin{abstract}
Strong lensing of gravitational waves is more likely for distant sources but predicted event rates are highly uncertain with many astrophysical origins proposed. Here we open a new avenue to estimate the event rate of strongly lensed systems by exploring the amplitude of the stochastic gravitational wave background (SGWB). This method can provide a direct upper bound on the high redshift binary coalescing rates, which can be translated into an upper bound on the expected rate of strongly lensed systems. We show that from the ongoing analysis of the Laser Interferometer Gravitational-wave Observatory (LIGO)-Virgo and in the future from the LIGO-Virgo design sensitivity stringent bounds on the lensing event rate can be imposed using the SGWB signal. Combining measurements of loud gravitational wave events with an unresolved stochastic background detection will improve estimates of the numbers of lensed events at high redshift. The proposed method is going to play a crucial in understanding the population of lensed and unlensed systems from gravitational wave observations.

\end{abstract}
\section{Introduction}
Gravitational waves from astrophysical sources are a new multi-frequency window to the Universe which ranges from the low-frequency regime ($\sim 10^{-9}$ Hz) originating from supermassive binary holes of masses $\mathcal{O}(10^9\, M_\odot)$ \footnote{The mass of the sun is denoted by $M_\odot\equiv 2 \times 10^{30}$ kg.} to high-frequency gravitational waves from binary neutron stars of masses ($\mathcal{O}(1 \, M_\odot)$). The nano-Hz gravitational wave signal can be measured using the International Pulsar Timing Array (IPTA) \citep{Arzoumanian:2018saf, Perera:2019sca, Hobbs:2017oam, 2010CQGra..27h4013H},    gravitational waves in the frequency range ($\sim 10^{-4}$ Hz- $10^{-1}$ Hz) will be probed using future space-based gravitational wave detector such as the Laser Interferometer Space Antenna (LISA) \citep{2017arXiv170200786A}, and the high-frequency gravitational waves in the frequency range ($\sim 10$ Hz- $10^{3}$ Hz) are measured from the network of ground-based gravitational wave detectors such as the currently ongoing HLO (Hanford Laser Interferometer Gravitational-wave Observatory), LLO (Livingston Laser Interferometer Gravitational-wave Observatory), VO (Virgo interferometer Observatory) \citep{Martynov:2016fzi, Acernese_2014}, and in the future from KAGRA (Kamioka Gravitational Wave Detector) \citep{Akutsu:2018axf}, and LIGO-India \citep{Unnikrishnan:2013qwa}. The proposed next-generation ground-based gravitational wave detectors such as Cosmic Explorer (CE) \citep{Reitze:2019iox} and Einstein Telescope (ET) \citep{Punturo:2010zz} are also expected to explore wider frequency range with lower instrument noise.

Gravitational wave signals can be broadly classified into two types, (i) high 
signal-to-noise ratio (SNR) individual gravitational waves events which we term "loud" events in the remaining of this paper \citep{LIGOScientific:2018mvr}, in contrast to unresolved gravitational wave sources contributing to the stochastic gravitational wave background (SGWB) as an incoherent contribution from a large number of coalescing binary events in the Cosmos \citep{Allen:1996vm, Maggiore:1999vm, Phinney:2001di, Regimbau:2007ed, Wu:2011ac, Rosado:2011kv, Zhu:2011bd,  Romano:2016dpx}. These two independent classes of gravitational wave signals are useful in combination for elucidating a vast range of possible astrophysical and cosmological origins over cosmic history, for which it is essential to distinguish the relative contributions of different astrophysics that may predominate.

One such cosmological effect which can shadow our understanding of the astrophysical processes is the level of strong gravitational lensing of gravitational waves generated by the cosmological structure in the foreground of gravitational wave sources \citep{PhysRevLett.80.1138, Wang:1996as,Dai:2016igl, Broadhurst:2018saj}. Gravitational lensing of gravitational waves will magnify the strain of the signal, which may make it possible to detect distant sources as individual loud events. If unrecognised, this magnification leads to a biassed underestimate of the true source distance and to  {an} overestimate of the chirp mass of binary events \citep{PhysRevLett.80.1138, Wang:1996as,Dai:2016igl,Broadhurst:2018saj, Contigiani:2020yyc}. Hence, our understanding of the nature of the source population of astrophysical binaries may be significantly affected, depending on the uncertain proportion of such lensed events\citep{Oguri:2018muv}. Apart from strong lensing of gravitational wave, the effect of weak lensing may also influence the strain statistics of gravitational waves, which can be explored to study several aspects of cosmology and to test the theory of gravity via the LIGO design sensitivity, and also from the future gravitational wave detectors such as LISA, ET, and CE \citep{Mukherjee:2019wcg, Mukherjee:2019wfw}. 

The impact of lensing on the derived properties of binary populations is potentially severe if the proportion of detected lensed events is large. The expected number of strongly lensed gravitational wave sources depends on two quantities, (i) properties of the population of gravitational lenses (ii) the coalescence rate of the GW binary sources and their redshift distribution. Several surveys of the strongly lensed systems of electromagnetic signals help in understanding the properties of strong lensing, in particular the complete far-infrared sky surveys where over 100 distant lensed galaxies now known \citep{2010ARA&A..48...87T}. Detailed lensing calculations have also been made, based on large-scale structure, that include CDM and baryons (stars) to estimate the expected lensing optical depth and the level of cosmic variance \citep{Hilbert:2007ny,2008MNRAS.386.1845H, Robertson:2020mfh}.  {However, in terms of the expected detection rate, the dominant source of uncertainty is the merger rate of the gravitational wave sources as the astrophysical origins are unclear (including field binaries and also binaries formed in the dense cores of young star clusters at high redshift \citep{1974CeMec..10..217H, 1974CeMec..10..185A, Portegies_Zwart_2002, 2010MNRAS.402..371B,2011EPJP..126...55D,Veske:2020zch}). From the O1+O2 observations of the LVC, there are no detections of any strongly lensed events \citep{2019ApJ...874L...2H}.}

We propose here a new avenue to limit the uncertainty in the lensing event rate by combining the measurements from the SGWB.  The binary merger rate and its redshift distribution can be probed using the SGWB signal, independently of the loud events, because the amplitude and shape of the power spectrum of SGWB energy density directly depend on the merger history of gravitational wave sources over cosmic history. As a result, we can infer the merger rates from the SGWB data, and use it to infer the expected numbers of loud lensed events. Even in the absence of a detection of the SGWB signal, this method can impose an upper bound on the number of lensed events. In this paper, we discuss this avenue and show the possible constraints on the lensed event rates which can be imposed using the measurements of the SGWB signal. The main findings of this paper are shown in Fig. \ref{nlf} and Fig. \ref{nlfd} using the publicly available LIGO O3 sensitivity and LIGO design sensitivity. Henceforth, any statement related to the LIGO O3 sensitivity and LIGO design sensitivity refers to the publicly available detector noise. \footnote{The noise curves are available publicly and can be downloaded from the following link \href{https://dcc-lho.ligo.org/LIGO-T2000012/public}{https://dcc-lho.ligo.org/LIGO-T2000012/public}}

The paper is organized as follows, in Sec. \ref{seclen} and Sec. \ref{secsgwb}, we set up the framework of strong gravitational lensing and SGWB signal respectively. In Sec. \ref{lensgwb}, we discuss the predictability of the lensed events using the SGWB signal. In Sec. \ref{bounds}, we show the possible bounds achievable from the observation of the LIGO-Virgo collaboration (LVC) on the SGWB signal and hence on the lensed event rates. We discuss the conclusion of this work and future scopes in Sec. \ref{conc}. 
\section{Strong Lensing of gravitational waves}\label{seclen}
Gravitational lensing of photons and gravitational waves in the presence of matter perturbations is an inevitable effect according to the general theory of relativity \citep{1992grle.book.....S, Bartelmann:2010fz}. In the geometric-optics limit, the phase of the gravitational waves remains unaltered, and the primary effect of gravitational lensing is evident in the strain of the signal,  and in the temporal and spatial aspects of the gravitational wave signal (see the review article \citep{Oguri:2019fix}).
 {Presence of matter perturbations between the source and observer leads to lensing of the gravitational waves which generates several observable effects such as (i) time-delay, (ii) magnification (demagnification), (iii) shift in the sky position, and (iv) multiple lensed images.}  Poor sky localization of the gravitational wave sources makes it impossible to measure the effect due to the shift in the sky position, However, the other four effects can be scrutinized from the gravitational wave signal. In this work, we explore the effect due to the magnification of gravitational waves. 

Magnification leads to change in the amplitude of the GW stain $h_{+, \, \times}(f_z)$ by the magnification factor $\mu$ (same for both the polarization states $'+, \, \times'$ )\citep{1987thyg.book.....H, Cutler:1994ys,Poisson:1995ef,maggiore2008gravitational}
\begin{align}\label{lens-1}
    \begin{split}
       h_{\pm}(f) (\hat n)= \sqrt{\mu}\sqrt{\frac{5}{96}}\frac{G^{5/6}\mathcal{M}_z^2 (f_z\mathcal{M}_z)^{-7/6}}{c^{3/2}\pi^{2/3}d_L}\mathcal{I}_{\pm} (\hat L.\hat n), 
           \end{split}
\end{align}
where $f_z= f/(1+z)$ is the redshifted frequency, $d_L$ is the luminosity distance to the gravitational wave source,  $\mathcal{I}_{\pm} (\hat L.\hat n)$ captures the projection of the angular momentum vector $\hat L$ on the line of sight $\hat n$, $\mathcal{M}_z= (1+z)\mathcal{M}$ is the redshifted chirp mass,  \footnote{Chirp mass $\mathcal{M}$ of a binary gravitational wave source is related to the mass of the individual compact objects ($m_1, m_2$) by the relation $\mathcal{M}= (m_1m_2)^{3/5}/(m_1+m_2)^{1/5}$.}. The redshifted chirp mass $\mathcal{M}_z$ of the gravitational wave sources can be well measured from the phased part of gravitational waves \citep{Cutler:1994ys}. But for the gravitational wave sources without electromagnetic counterparts, redshift cannot be  independently estimated. As a result, the degeneracy between source chirp mass and redshift cannot be lifted. 

The main effect of lensing magnification leads to a wrong inference of the luminosity distance to the source $\tilde{d_L}=d_L/\sqrt{\mu}$. 
This implies that for $\mu >1$, the inferred luminosity distance $\tilde{d_L}$ to the gravitational wave sources will be smaller than the actual luminosity distance $d_L$. As a result, from the inference of the redshift using luminosity distance $\tilde{d_L}$ and the best-fit cosmological parameters \citep{Aghanim:2018eyx}, we can estimate the biased cosmological redshift $\tilde{z}$ which is going to be smaller than the true  cosmological redshift $z_s$. As a result, the estimated source chirp mass $\mathcal{M}$ will be biased towards a higher value $\tilde{\mathcal{M}}\equiv \mathcal{M}_z/(1+\tilde{z}) = \mathcal{M} (1+z)/(1+\tilde{z}) > \mathcal{M}$.\footnote{This will also lead to a biased estimate of the individual source masses as $\tilde{m}_i\equiv m_{i,z}/(1+\tilde{z})= m_i (1+z)/(1+\tilde{z})> m_i$.} This implies that the impact of strong lensing can lead to a biased measurement of the source parameters of the gravitational wave sources. This can be an obstacle for correctly inferring the gravitational wave source properties from a sample of objects if the number of strongly lensed events is large.

The number of detectable lensed events of magnification factor $\mu$ can be written in terms of the spatial distribution of the structures along the line of sight, redshift of the source $z_s$, gravitational wave source parameters such as mass, spin, which are denoted by $\theta$, merger rate of the gravitational wave sources $R(z_s, \theta)$ having source parameters $\theta$, and the gravitational wave detector detector response function $\mathcal{S}(\theta,z_s, \mu)$ 
\begin{align}\label{nzl-1}
    \begin{split}
     \dot N_l (\geq \mu) \equiv  \frac{dN}{dt} (\geq \mu, z_s)= \int_0^{z_s}dz \int d\theta & \overbrace{\frac{dV}{dz} \tau_l(\geq \mu, z)}^{\text{Cosmology}} \overbrace{p(\theta) \frac{\mathcal{R}(z, \theta)}{(1+z)}}^{\text{Astrophysics}}\\ &\times \overbrace{\mathcal{S }(\theta,z, \mu)}^{\text{Detector response}} ,\\
           \end{split}
\end{align}

The above equation has three kinds of terms, the detector response part, the astrophysical part, and the cosmological part. The number of detected lensed events of magnification factor $\mu$ depends on the redshift dependence of all of  these terms. We briefly explain below each of these terms and their redshift dependence. 

\textit{Detector response part:} The detectability of the gravitational wave signal $h^{obs}(f)= \sum_{\pm}F_{\pm}(\alpha, \beta, \i)h_{\pm}(f)$, expressed in terms of the antenna function $F_{\pm}(\theta, \phi, \i)$ and signal $h_{\pm}$ (given in Eq. \eqref{lens-1}) depends on the matched-filtering 
signal-to-noise ratio (SNR) $\rho$ which can be defined as \citep{Sathyaprakash:1991mt, Cutler:1994ys,Balasubramanian:1995bm}
\begin{equation}\label{snrgw}
    \rho^2\equiv 4Re\bigg[\int_0^{f_{max}} df \frac{ h^{obs}(f)g^*(f)}{S_n(f)}\bigg],
\end{equation}
where $g(f)$ denotes the waveform of the gravitational wave signal which is used in the matched-filtering method and $S_n(f)$ is the noise power spectrum.  We choose the value of $f_{max}= f_{merg}\equiv c^3(a_1\eta^2+a_2\eta+a_3) /(G\pi M)$  
in this analysis \citep{Ajith:2007kx}\footnote{The values of the coefficients $a_1$, $a_2$, and $a_3$ are given in the table \ref{tab:params}.}, where $M=m_1+m_2$ is the total mass of the binary system and $\eta= m_1m_2/M^2$ is the symmetric mass ratio. For a fixed gravitational wave chirp mass $\mathcal{M}_c$, the detectability of a signal varies, (to first order),  with the value of $\sqrt{\mu}/d_L$. As a result, the SNR of detection of an event decreases with an increase in the luminosity distance, and for events below a chosen detection threshold $\rho_{th}=8$, we cannot make a statistically significant detection as a loud event. We define detector response function as Heaviside step function $\mathcal{S}(\theta,z_s, \mu)\equiv H (\rho (\mu, d_L, \mathcal{M}_c)-\rho_{th})$ \footnote{Heaviside step function $H(x)=1$, only when the argument satisfies the criterion $x\geq 1$.} which assures that only the gravitational wave sources for which $\rho (\mu, d_L, \mathcal{M}_c) \geq 8$, can be detected as individual events.
\begin{figure*}
\centering
\includegraphics[trim={0cm 0.cm 2cm 2.cm},clip,width=0.8\textwidth]{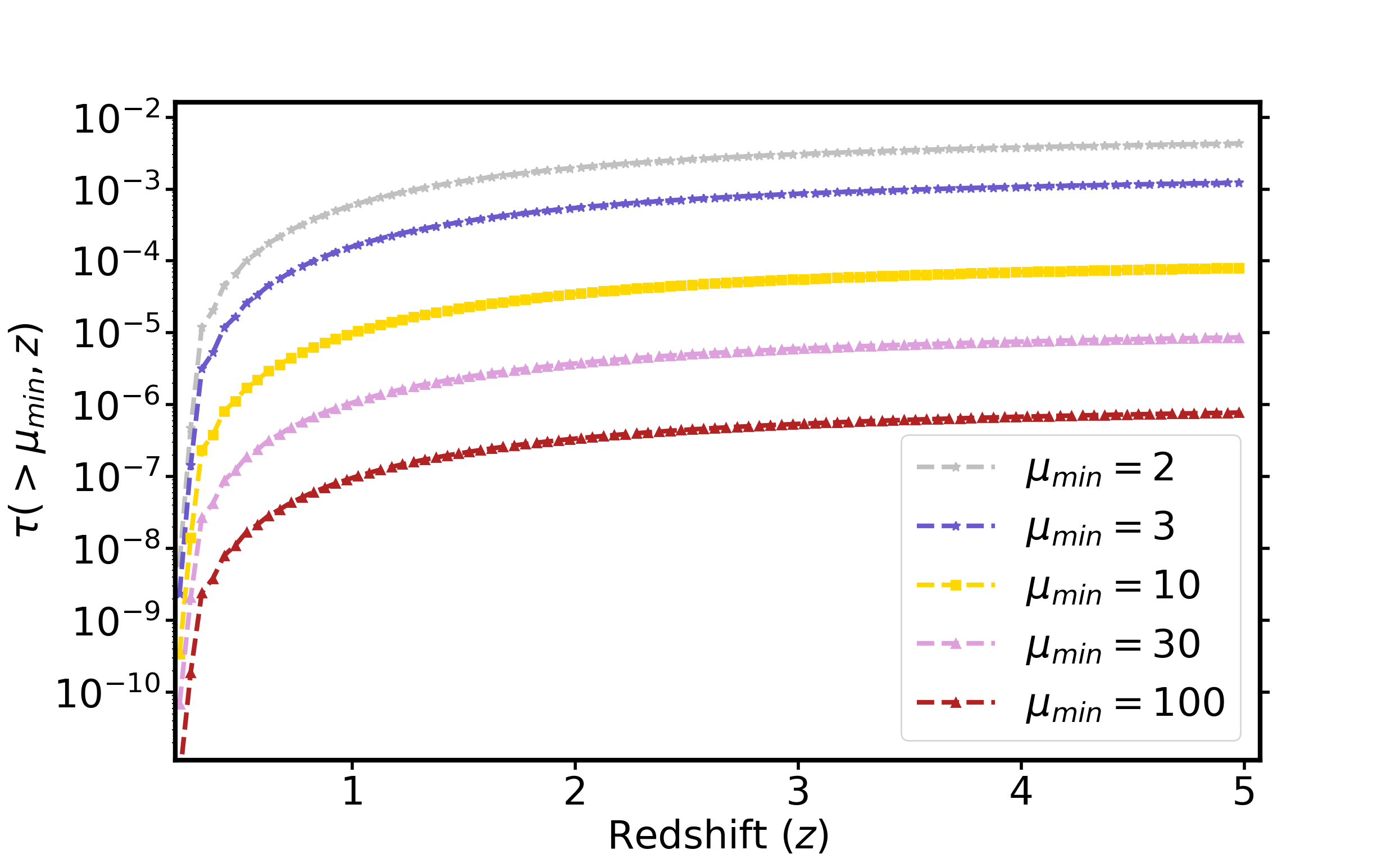}
\caption{Lensing optical depth $\tau(>\mu, z)$ 
shown for different magnification factors $\mu$ as a function of cosmological redshift. The plots indicates that the lensing optical depth is an increasing function of redshift, and is suppressed for highly magnified systems $\mu>>1$.}
\label{Fig:tau}
\end{figure*}

\textit{Astrophysical part:} This part consists of two terms, $p(\theta)$ which is the probability distribution of the gravitational wave source parameters such as the masses of the compact objects $m_i$, spin $\chi_i$, and $\mathcal{R}(z,\theta)$ is the event rate of the gravitational wave sources, with parameters $\theta$, per comoving volume in the source frame at redshift $z$. Both these quantities are not yet well-known \citep{LIGOScientific:2018mvr}. The mass distribution of the binary mass components are usually considered to follow a power-law of the form $1/m_i^{2.35}$ or flat in log-space $1/m_i$. Sources detected by LVC from the O1 and O2 data indicates a local redshift event rate of binary black hole (BBH) mergers about $10^2$ Gpc$^{-3}$ yr$^{-1}$ \citep{LIGOScientific:2018mvr}. The event rate at high redshift and its evolution with redshift is poorly constrained. We can expect the merger rates to follow the star formation rate at high redshift. If the merger rate is a decreasing function of redshift, and consistent with the local event rate measured by LVC \citep{Kalogera:2006uj, LIGOScientific:2018mvr}, then the product of the lensing optical depth and merger rates is going to be negligible. On the other side, the scenarios for which the merger rate is constant or increases with redshift can produce non-negligible lensed events. 
\begin{figure*}
\centering
\includegraphics[trim={0cm 0.cm 2.7cm 0.cm},clip,width=0.8\textwidth]{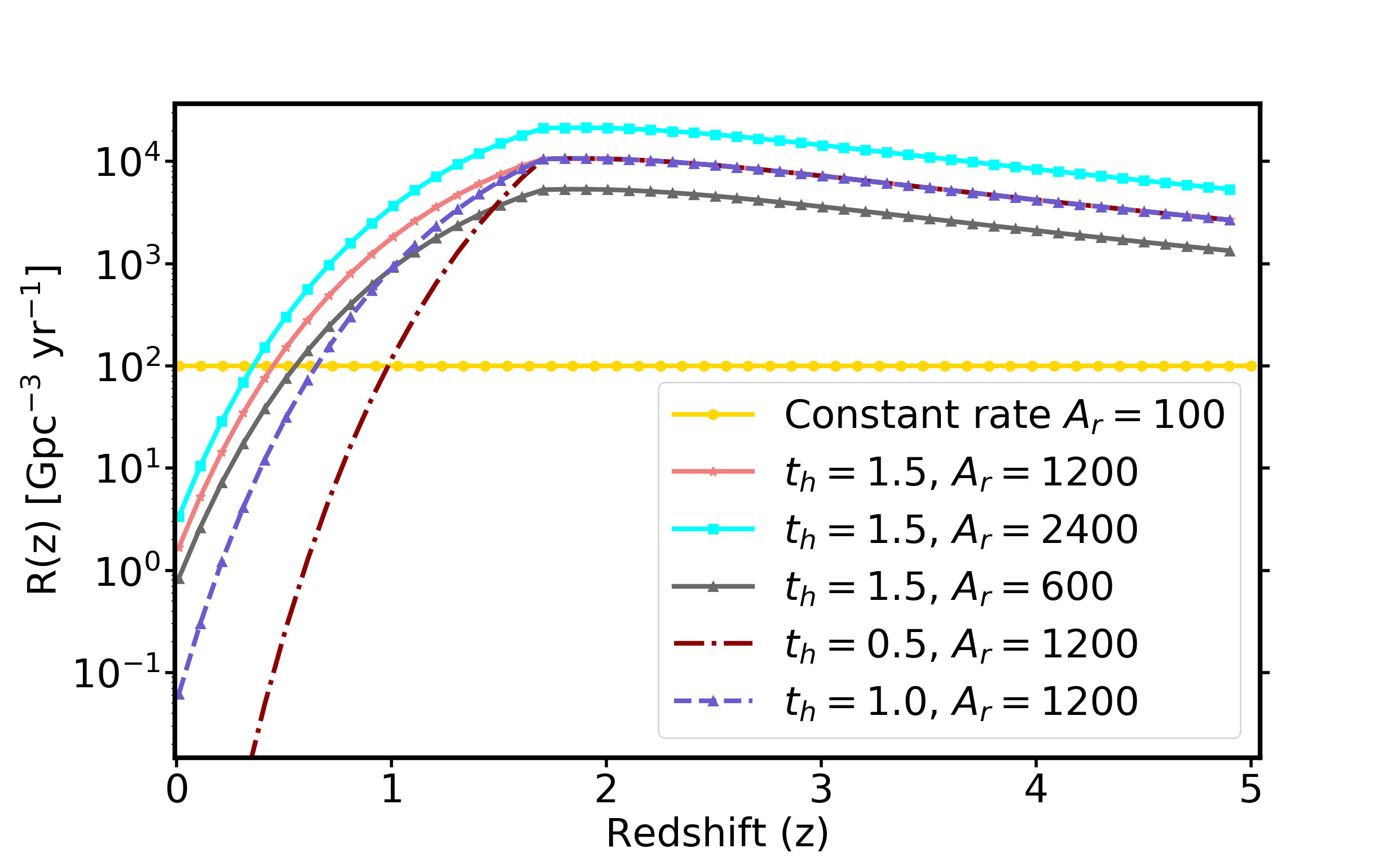}
\caption{The merger rates of BBH sources as a function of the cosmological redshift are shown for different choices of amplitude $A_r$ (in units of Gpc$^{-3}$ yr$^{-1}$), and the half-life $t_h$ (in units of Gyr). }
\label{rates}
\end{figure*}

\textit{Cosmological  part:} This part consists of the cosmological volume element which can be written in terms of the comoving distance $d_c$ as  $\frac{dV}{dz}= \frac{4\pi c d_c^2}{H(z)}$, and the lensing optical depth $\tau_l(\geq \mu, z)$ can be written as \citep{Turner:1984ch}
\begin{align}\label{nzl-2}
    \begin{split}
      \tau_l(\geq \mu, z)= \int_0^z dz_l \frac{d\tau(\geq \mu, z, z_l)}{dz_l},
           \end{split}
\end{align}
 {where, $d\tau(\geq \mu, z, z_l) / dz_l$  
is the differential lensing optical depth which can be written as \citep{Turner:1984ch}
\begin{align}\label{nzl-2a}
    \begin{split}
     \frac{d\tau(\geq \mu, z, z_l)}{dz_l}= \frac{1}{A_T(z_s)} \frac{dV(z_l)}{dz_l}\int dM \frac{dN}{dMdz} A_N(\mu, M, z_l, z_s),
           \end{split}
\end{align}
where, $A_T(z_s)$ is the area of the spherical shell at redshift $z_s$ in the physical units, $\frac{dV(z_l)}{dz_l}$ is the differential volume fraction at the redshift of the lens $z_l$, $\frac{dN}{dMdz}$ is the halo mass function per unit halo mass and redshift, and $A_N(\mu, M, z_l, z_s)$ is the area for magnification higher than $\mu$ computed in the image plane, but divided by the factor $\mu$ to account for the equivalent area in the source plane.}   

In the absence of any information of the gravitational wave redshift, the integration over redshift can extend up to some value $z_{max}$. We show the lensing optical depth as a function of the source redshift $z_s$ for different magnification factor in Fig. \ref{Fig:tau} using the model as described by \citep{Watson:2013mea,Diego:2018fzr, Diego:2019rzc, Diego2020}. For the optical depth calculation, we have considered the mass range from $10^{11}\, M_{\odot}$ to $3\times 10^{15}\, M_{\odot}$ for the mass function given by \citet{Watson:2013mea}.  {As described in \cite{Diego2020}, each halo is modelled as an elliptical NFW with an additional central cusp component to account for baryons. This baryonic component is particularly important in relatively small halos, that can become supercritical around the central baryonic overdensity.}
The plot indicates that the lensing optical depth decreases with increase in magnification factor $\mu$ and is larger for sources at high cosmological redshifts than at low redshifts. As a result, high magnification for a low redshift source is less probable than a system at high redshift. In particular, baryons can play a significant role especially in small halos that may become supercritical if their cores are dense enough thanks to baryon cooling. Also, the slope of the potential around the critical curves plays an important role in the resulting magnification near the critical curves, hence profiles like NFW or SIS may predict different optical depths. Finally, substructure near the critical curves may promote smaller magnification factors (tens to hundreds) at the expense of more extreme values (thousands). A recent study by \citet{Robertson:2020mfh} has calculated the variation of the lensing optical depth from simulations with redshift and masses \footnote{See Fig. 5 in \citet{Robertson:2020mfh}}. This shows the agreement between different models and simulations for the lensing optical depth $\tau (\mu >10, z)$ and the typical variations depending upon the modelling. The lensing optical depth estimation considered in this work is also consistent with these results, except for the case considered by \citep{2019ApJ...874L...2H} which has considered larger values than what is  expected from currently known models and simulations.

To explore the range of variation in the number of lensed events, we consider the case with constant merger rate $A_r=100$ Gpc$^{-3}$ yr$^{-1}$, motivated from the current measurements with gravitational wave observations \citep{LIGOScientific:2018mvr}.  {We consider this as the benchmark model in our analysis.} Along with this model, we also consider a few other models of BBH merger rates which match with the redshift distribution of star formation rate \citep{Madau:2014bja} in high redshift galaxies, and can produce larger number of lensed events than the case with constant merger rates. \citep{Broadhurst:2018saj, Broadhurst:2019ijv, Diego:2019rzc, Broadhurst:2020moy}. The main reason to consider this model is because it has been used to claim that several of the detections made by the LVC in the O1 and O2 runs, are already strongly magnified events \citep{Broadhurst:2018saj, Broadhurst:2019ijv, Broadhurst:2020moy}, and it is interesting to see what kind of predictions these types of models make for the SGWB signal.  {In order to understand the relation between the SGWB amplitude and lensing event rates, we consider a wide range of models with large event rates at high redshift which are allowed from the SGWB observations from O1 and O2. For the benchmark model, we can expect a few $\times 10^3$ BBH mergers per year with the design sensitivity of LVK.}
We consider a parametric model of the merger rates, with two free parameters (i) the amplitude of the merger rates denoted by $A_r$, and the half-life of the exponential decay which is denoted by ${t}_h$
 \begin{align}\label{merger-1}
    \begin{split}
        R(z_s,\theta)= &\frac{A_r(1.0+z_s)^{2.7}}{(1.0 + ( \frac{(1.0+z_s)}{2.9})^{5.6})} \\&  \times 
        \begin{cases}
   \exp\bigg(\frac{-|({t}_{z_s}-{t}_{z=z_p})|}{({t}_h)}\bigg),& \text{if } z_s< z_p,\\
    1,              & \text{otherwise}.
\end{cases}
   \end{split}
\end{align}  
In the above equation, ${t}_{z_i}$ denotes the look-back time to redshift $z_i$, $z_p=1.7$ denotes the pivot redshift below which the exponential decay is important, and higher than which the merger rates follow the redshift distribution of the star formation rate. We plot the event rates for different values of the parameter $A_r$, and $t_h$ in Fig. \ref{rates}.

\begin{figure*}
\centering
\includegraphics[trim={0cm 0.cm 2.7cm 0.cm},clip,width=0.8\textwidth]{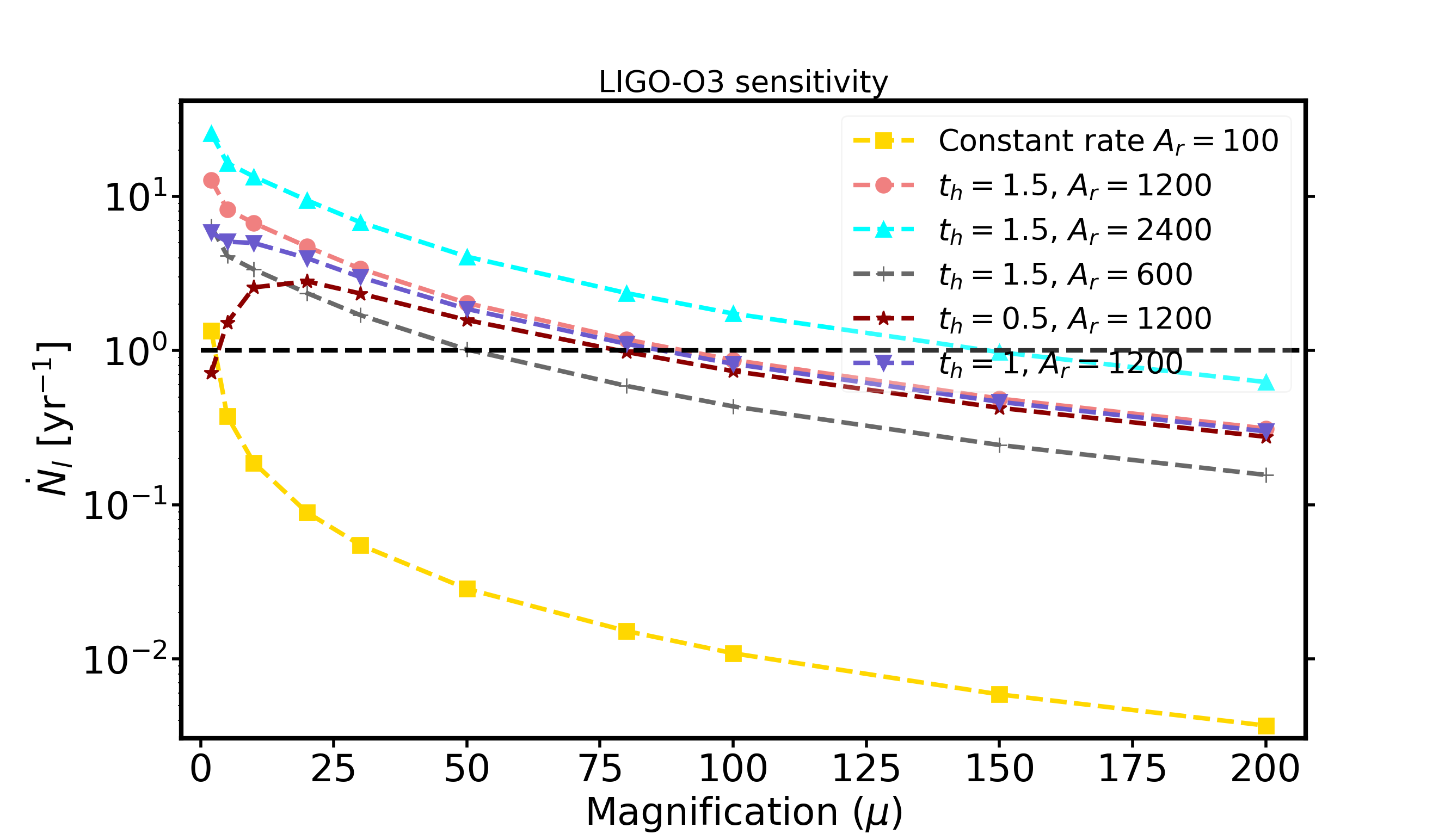}
\caption{We show the detectable lensed event rates as a function of magnification factor $\mu$ for different merger rates $R(z)$ of BBHs for the LIGO O3 sensitivity. The redshift dependences of the merger rates $R(z)$ are shown in Fig. \ref{rates} with the amplitude $A_r$ (in units of Gpc$^{-3}$ yr$^{-1}$), and the half-life $t_h$ (in units of Gyr). The lensing event rate is obtained by integrating Eq. \ref{nzl-1} up to $z_s=5$. The black dashed line denotes $\dot N_l=1$.}
\label{lensedratesO3}
\end{figure*}

\begin{figure*}
\centering
\includegraphics[trim={0cm 0.cm 2.7cm 0.cm},clip,width=0.8\textwidth]{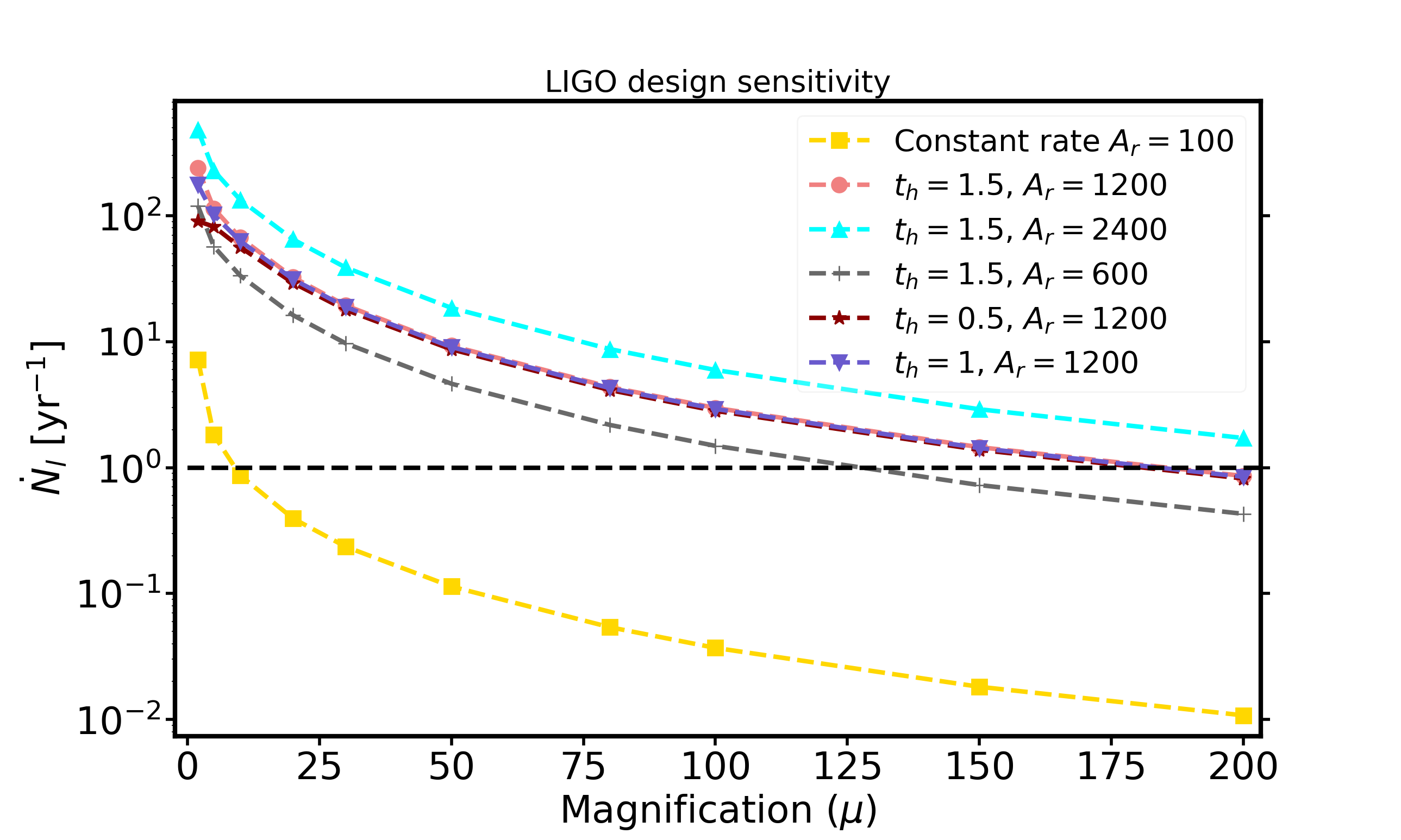}
\caption{We show the detectable lensed event rates as a function of magnification factor $\mu$ for different merger rates $R(z)$ of BBHs using the advanced LIGO design sensitivity with the amplitude $A_r$ (in units of Gpc$^{-3}$ yr$^{-1}$), and the half-life $t_h$ (in units of Gyr). The redshift dependence of the merger rates $R(z)$ is shown in Fig. \ref{rates}. The lensing event rate is obtained by integrating Eq. \ref{nzl-1} up to $z_s=5$. The black dashed line denotes $\dot N_l=1$.}
\label{lensedratesdesign}
\end{figure*}

Using the instrument noise corresponding to the LIGO O3 sensitivity, and the design sensitivity, we estimate the expected number of lensed events per earth year $\dot N_l$ for different merger rates and magnification factor $\mu$ in Fig. \ref{lensedratesO3} and Fig. \ref{lensedratesdesign} respectively.  The distribution function of the individual black hole masses are considered as $p(\theta)=1/m^{2.35}$ $\theta \in \{m_1, m_2\}$ with the mass range of black holes $m_i \in [5 M_\odot, 50 M_\odot]$, and the maximum source redshift of the BBH mergers are considered up to $z_s=5$.  {In this analysis, we have integrated over the orientation function $\Theta$ of the GW sources, assuming the probability distribution $P(\Theta)= 5\Theta(4-\Theta)^3/256$ for $\Theta \in [0,4]$ \citep{Finn:1992xs, Finn:1995ah,Ng:2017yiu}. For sources with $\Theta \approx 4$, one can observe up to high redshift, and for the GW sources with $\Theta \approx 0$ can only be detected only up to very low redshift.} For the O3 sensitivity and design sensitivity of  LIGO detectors, we show the lensing event rates in Fig. \ref{lensedratesO3} and Fig. \ref{lensedratesdesign} respectively. The number of lensed events $\dot N_l$ varies due to the variation in the merger rate and also due to the redshift reach of the gravitational wave detectors. Cases with higher event rates at high redshift lead to a higher number of lensed events than for the case with a constant event rate. The number of lensed events for higher magnification factor is also suppressed due to the smaller value of the lensing optical depth (as shown in Fig. \ref{Fig:tau}). The contribution to the detectable lensed events with magnification factor $\mu$ depends on the detector response function $\mathcal{S}(\theta, \mu, z)$ which goes to zero for the events below the detection threshold. As a result, lensed events with smaller magnification factors are detectable from low redshift, and vice-versa. This aspect is manifested in the lensed event rates for the case shown by the brown line ($A_r= 1200$ Gpc$^{-3}$ yr$^{-1}$, $t_h= 0.5$ Gyr) and yellow line (constant rate $A_r= 100$ Gpc$^{-3}$ yr$^{-1}$). The case shown by the yellow line has larger merger rate $R(z)$ at low redshift than for the case shown by the brown line (see Fig. \ref{rates}) for $z<0.98$. As a result, the events with low magnification $\mu \leq 5$, have a larger number of lensed events for the yellow line, than the brown line. For the magnification factor $\mu \geq10$, lensed events are arising primarily from the high redshift, for which the merger event rate $R(z)$ is larger for the brown line than the yellow line (see Fig. \ref{rates}). 

Fig. \ref{lensedratesO3} and Fig. \ref{lensedratesdesign} manifests that the lensed event rates can vary by a few orders of magnitude for different magnification factor depending upon the merger rate $R(z)$. This uncertainty associated with the lensed event rates due to the uncertainty in the merger rate predominates over the expected uncertainty in the lensing optical depth estimation, particularly for magnification factor $\mu>>1$. As a result, the possibility of inferring the merger rates at high redshift is going to be useful to limit the uncertainty in lensed event rates. A possible bound on the expected lensed events will also be useful for understanding its impact on the population of gravitational wave sources.   

\begin{table}
    \centering
\begin{tabular}{||p{1.5cm}|p{1.5cm}|p{1.5cm}|p{1.5cm}|}
\hline
         $X_i$ &$a_1$ ($\times 10^{-1}$) &$a_2$ ($\times 10^{-2}$) &$a_3$ ($\times 10^{-2}$)\\
         \hline
         \hline
           $f_{merg}$ & $2.9740$ & $4.4810 $ & $9.5560$\\
          $f_{ring}$ & $5.9411$ & $8.9794$ & $19.111$\\
          $f_{cut}$ & $8.4845$ & $12.848$ & $27.299$\\
          $f_{w}$& $5.0801$ & $7.7515$ & $2.2369$\\
          \hline
    \end{tabular}
    \caption{We show the values of the parameters to obtain the frequency $f_{merg}$, $f_{ring}$, $f_{cut}$, and $f_w$ denoted by the functional form $X_i= c^3(a_1\eta^2 + a_2\eta +a_3)/\pi G M$ \citep{Ajith:2007kx}.}
    \label{tab:params}
\end{table}

\section{Stochastic gravitational wave background}\label{secsgwb}
The merger rate of the astrophysical gravitational wave sources can also be detected using the stochastic gravitational wave background \citep{Allen:1996vm, Maggiore:1999vm, Phinney:2001di, Wu:2011ac, Regimbau:2007ed, Romano:2016dpx, Rosado:2011kv, Zhu:2011bd, 10.1093/mnras/stz3226, Boco:2019teq, Callister:2020arv}. The combined contribution of gravitational wave emission from all the compact objects within the horizon leads to an astrophysical stochastic gravitational wave background which can be written in terms of the GW merger rates $R(z,\theta)$ in a comoving volume $\frac{dV}{dz}$ between the cosmic time $t(z)$ and $t(z+\Delta z)$ by the relation
\citep{Phinney:2001di, Regimbau:2007ed,Zhu:2011bd}
\begin{align}\label{sgwb-1}
    \begin{split}
        \Omega_{GW} (f)= \frac{f}{\rho_cc^2}\int_0^\infty dz & \int d\theta  \overbrace{\frac{dV}{dz}}^{cosmology}\overbrace{p(\theta)\frac{R(z, \theta)}{(1+z)}}^{astrophysics} \\& \times \overbrace{\bigg(\frac{1+z}{4\pi c d_L^2}\frac{dE_{GW} (\theta)}{d f_r}\bigg)}^{GW source}\bigg|_{f_r= (1+z)f}, 
    \end{split}
\end{align}
where, $p(\theta)$ (as discussed in Sec. \ref{seclen}) is the probability distribution of the source parameters such as source masses $m_i$, spin $\chi_i$. The terms $p(\theta)$ and $R(z, \theta)$ appearing in the theoretical estimate of the SGWB signal (given in Eq. \eqref{sgwb-1}) are the same as the one which appears in the estimation of the lensed event rate (given in Eq. \ref{nzl-1}). $\frac{dE_{GW}}{d f_r} (\theta)$ is the energy emission per frequency bin in the source frame which can be written in terms of the chirp mass $\mathcal{M}_c$ of the gravitational wave sources as
\begin{align}\label{sgwb-1a}
    \begin{split}
        \frac{dE_{GW}(\theta)}{d f_r}= \frac{(G\pi)^{2/3}\mathcal{M}_c^{5/3}}{3} \mathcal{G}(f_r), 
    \end{split}
\end{align}
where $\mathcal{G}(f_r)$ captures the frequency dependence during the inspiral, merger, and ringdown, phase of the gravitational wave signal \citep{Ajith:2007kx}
\begin{equation}\label{fr-dep-1}
    \begin{split}
        \mathcal{G}(f_r)=
        \begin{cases}
        f_{r}^{-1/3} \, \text{for}\, f_r < f_{merg},\\  
         \frac{f_{r}^{2/3}}{f_{merg}}\, \text{for} \, f_{merg} \leq f_r < f_{ring},\\
         \frac{1}{f_{merg}f^{4/3}_{ring}}\bigg(\frac{f_{r}}{1+(\frac{f_r-f_{ring}}{f_w/2})^2}\bigg)^2\, \text{for}\,  f_{ring} \leq f_r < f_{cut},
         \end{cases}
    \end{split}
\end{equation}
where the values of $f_{x}$ can be expressed in terms of the polynomial relation $c^3(a_1\eta^2 + a_2\eta +a_3)/\pi G M$ in terms of total mass $M= m_1 + m_2$ and symmetric mass ratio $\eta= m_1m_2/M^2$. The values of $a_1, a_2$, and $a_3$ for different $f_x$ are mentioned in table \ref{tab:params} \citep{Ajith:2007kx}. For  given coalescing binaries of masses $m_1$ and $m_2$, the binaries will be emitting gravitational waves in the inspiral part up to frequency $f_{merg}$, followed by the ringdown part up to frequency $f_{ring}$, and will stop  emitting gravitational wave signal after $f_{cut}$. $f_w$ denotes the width of the Lorentzian function. 

\begin{figure*}
\centering
\includegraphics[trim={0cm 0.cm 2.7cm 0.5cm},clip,width=.8\textwidth]{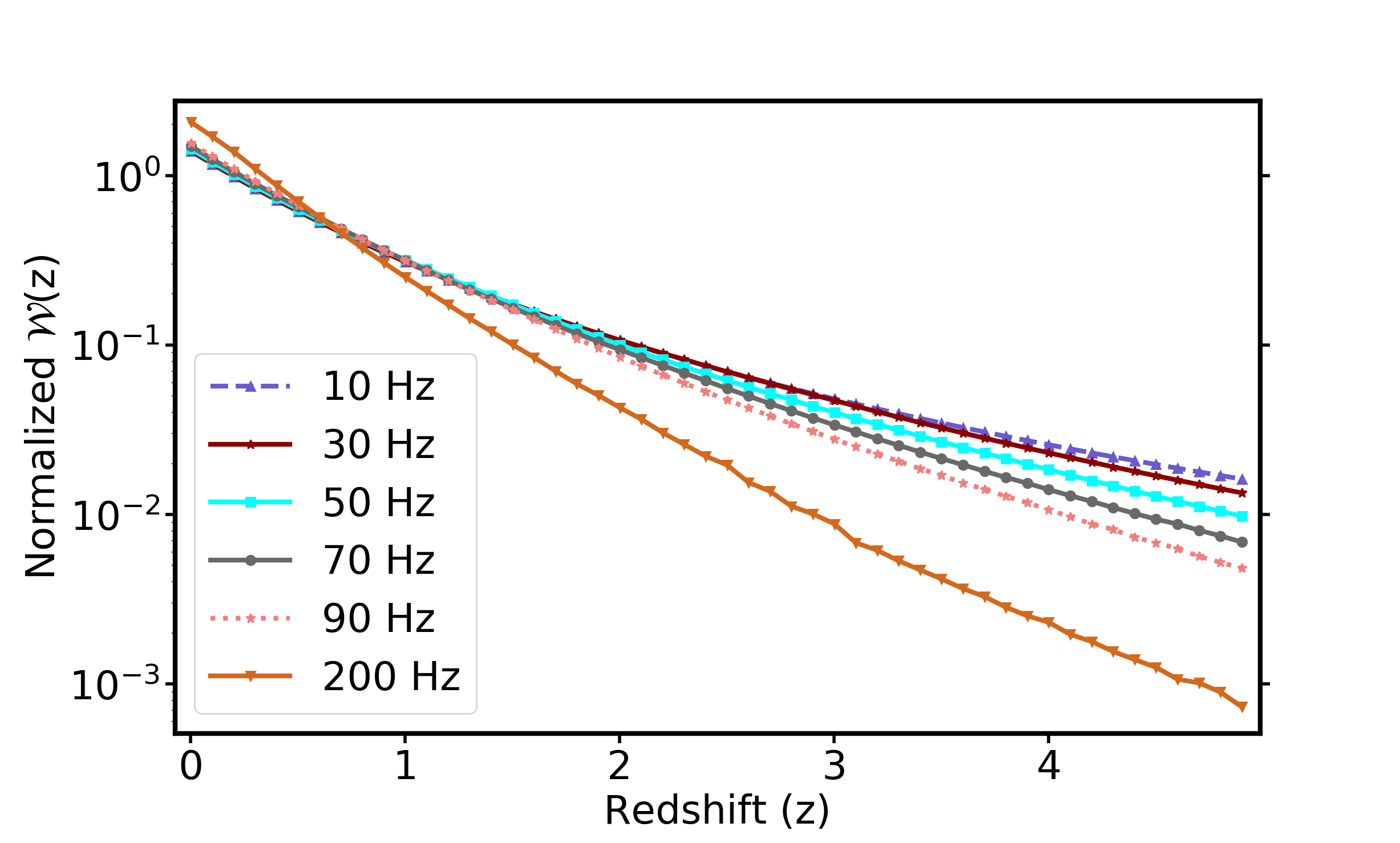}
\caption{The normalized window function of the SGWB signal for different frequencies are shown as a function of the redshift $z$. This plots shows that the low frequency part of the SGWB signal is a better probe of the high redshift event rate than the high frequency SGWB signal.}
\label{window}
\end{figure*}

Most of the signal in the SGWB arises from the inspiral part during which the system spends a longer time, compared to the merger and the ringdown part. As the maximum frequency of emission of the gravitational wave emission is inversely proportional to the total mass of the compact objects $M$, so the contribution to the low frequency part of the SGWB signal comes from the high masses, and  lighter masses will emit up to a larger frequency. Also due to cosmological redshift, gravitational wave source emitting at a frequency $f_r$ will be observed at a smaller frequency $f= f_r/(1+z)$. This implies that the low frequency part of the observed SGWB signal arises from redshifted high mass objects which are coalescing in the observable Universe, and high frequency SGWB signal gets its contribution from the redshifted lighter masses. The distribution of the events of the SGWB is expected to follow a Poisson distribution according to the event rate $R(z, \theta)$. The Poisson nature of the gravitational wave sources leads to temporal fluctuations on the SGWB which depends on the events and the type of the gravitational wave sources (BNS, NS-BH, or BBHs)  \citep{10.1093/mnras/stz3226}. In this analysis, we consider the mean value of the SGWB signal and will not explore its temporal fluctuations. 
 
Though the SGWB is an integrated contribution up to high redshift, we can define the window function of the SGWB signal for different frequency  $\mathcal{W}(z,f)$ as 
 \begin{align}\label{sgwb-3}
    \begin{split}
        \Omega_{GW} (f)= \frac{f}{\rho_c c^2}\int_0^{z_{max}} dz\, \mathcal{W}(z,f) R(z).
    \end{split}
\end{align}
 The window function $\mathcal{W}(z,f)$  can manifest the relative contributions to the SGWB signal from different redshifts and frequencies. By considering the distribution function of the gravitational wave source parameters in the mass range $[5\,M_\odot, 50\,M_\odot]$ as $p(\theta) = 1/m_1^{2.35}$ (same as the one the one considered for estimation of the strong lensing event rates in Sec. \ref{seclen}) and the redshift distribution of gravitational wave merger rates mentioned in Eq. \eqref{merger-1} (and shown in Fig. \ref{rates}), we show the normalized window function $\mathcal{W}(z,f)\equiv \int_0^{z_{max}} dz W(z, f,)=1$ in Fig. \ref{window}.  The maximum source redshift $z_{max}=5$ of the gravitational wave sources is considered in Eq. \eqref{sgwb-3}. For this estimation we do not consider the contribution from non-zero spin parameters. Non-zero spin parameters can cause only mild variations in the amplitude of the SGWB signal for $f<100$ Hz \citep{Zhu:2011bd}. So, our conclusion in the analysis will not change by the inclusion of spin, and can also be included for future joint study of the lensing event rate and SGWB signal. 
 
 The window function $\mathcal{W}(z,f)$ shows that the relative contribution from the low redshift Universe is larger for high frequencies, and the relative contribution from the high redshift gravitational wave mergers is mostly from the low-frequency gravitational wave signal. This arises because the mergers at high redshift contribute more at  low frequencies due to the cosmological redshift of the gravitational wave frequency $f= f_r/(1+z)$. The cross-over between the relative contributions between high frequency and low frequency happens around $\sim z=0.7$, which depends on the distribution function of the gravitational wave source parameters $p(\theta)$. The window function $\mathcal{W}(z,f)$ indicates that the measurement of the low-frequency SGWB signal is a better probe of the high redshift mergers than the high-frequency gravitational wave signal.
 
 \begin{figure*}
\centering
\includegraphics[trim={0cm 0.cm 2.cm 0.5cm},clip,width=.8\textwidth]{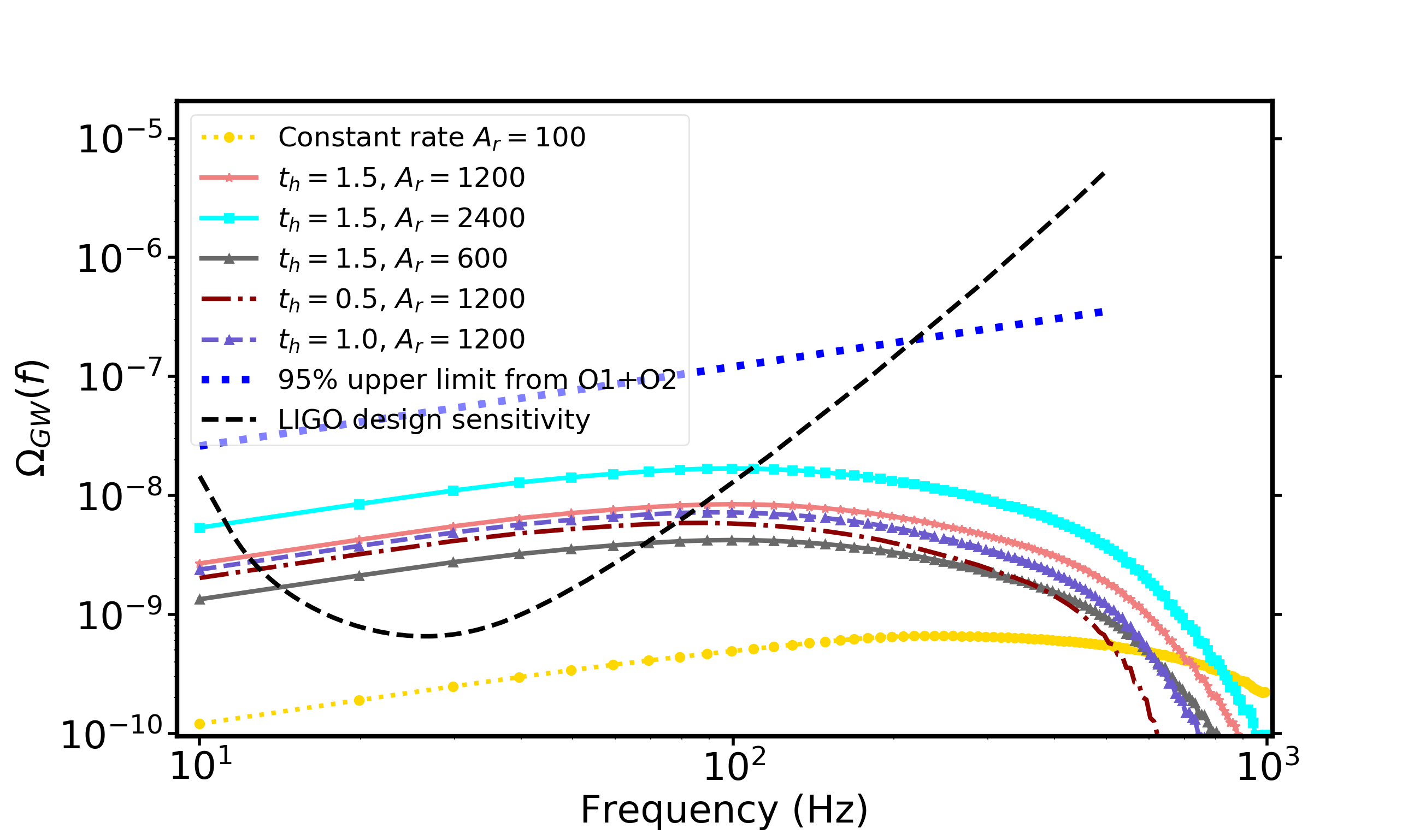}
\caption{We show the power spectrum of the SGWB signal as a function of frequency for different merger rates of the BBHs in the mass range $[5\, M_\odot, 50\, M_\odot]$. The mass distribution of each of the merging black holes is considered as a power-law ($p(\theta)= 1/m^{2.35}$ for both of the masses). The maximum redshift of the gravitational wave sources is considered up to $z_{max}=5$.  {We have also shown the $95\%$ upper limit on the SGWB power spectrum from the O1+O2 observation by the blue dotted line \citep{LIGOScientific:2019vic} and the}  $1$-$\sigma$ power-law integrated (PI) curve for the LIGO design sensitivity with two years of integrated time in dashed black line for the frequency dependence $f^{2/3}$ \citep{TheLIGOScientific:2016wyq}.} 
\label{sgwball}
\end{figure*} 

\begin{figure*}
\centering
\includegraphics[trim={0cm 0.cm 2.cm 0.5cm},clip,width=.8\textwidth]{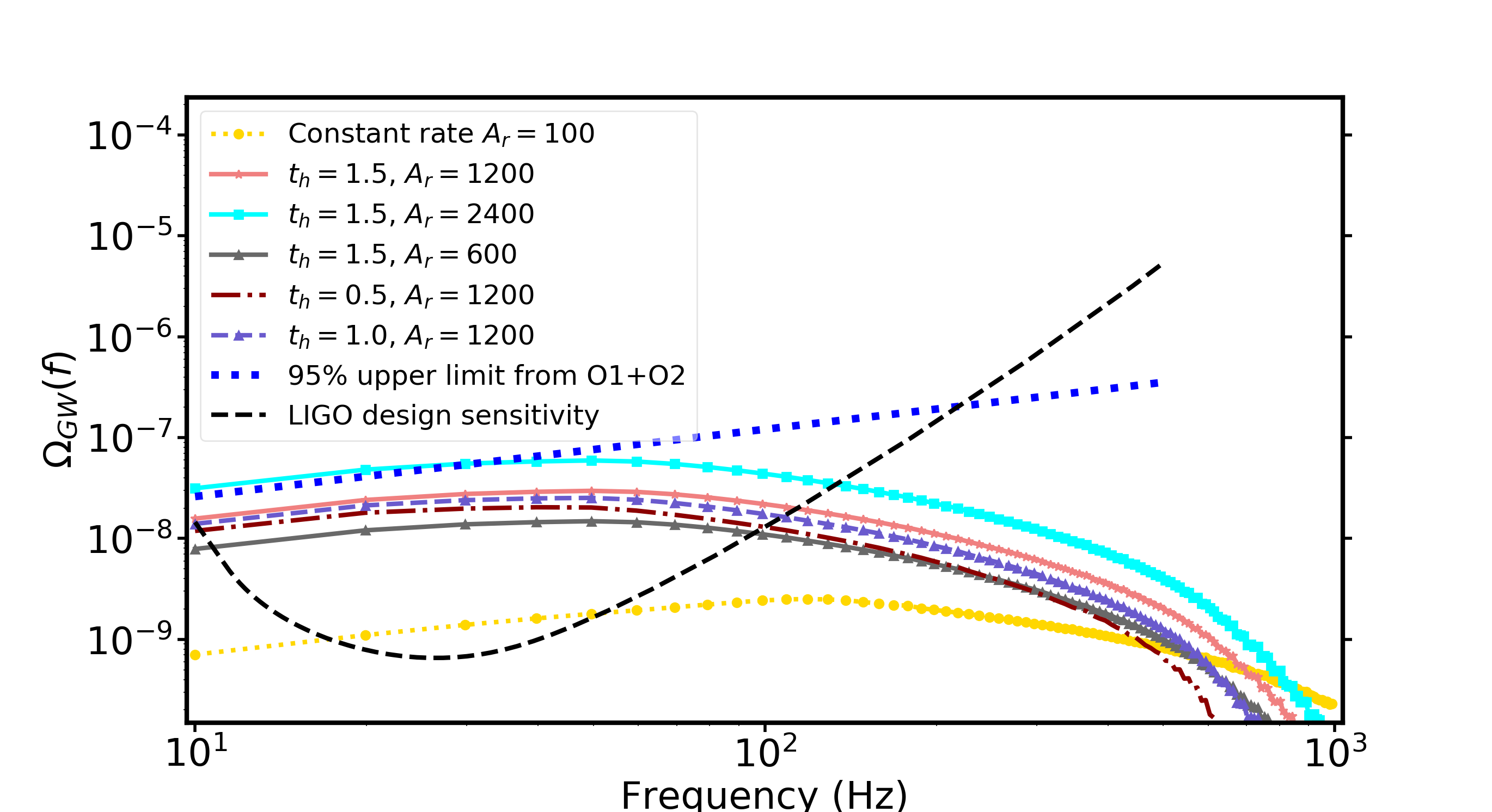}
\caption{We show the power spectrum of the SGWB signal as a function of frequency for different merger rates of the BBHs in the mass range $[5\,M_\odot, 95\,M_\odot]$. The mass distribution of each of the black holes is considered as a power-law ($p(\theta)= 1/m^{2.35}$ for the heavier object) and flat-log $p(\theta)= 1/m$ for the lighter mass. The maximum redshifts of the gravitational wave sources are considered up to $z_{max}=5$. The redshift distribution of the merger rates is shown in Fig. \ref{rates} with the amplitude $A_r$ (in units of Gpc$^{-3}$ yr$^{-1}$), and the half-life $t_h$ (in units of Gyr).  {We have also shown the $95\%$ upper limit on the SGWB power spectrum from the O1+O2 observation by the blue dotted line \citep{LIGOScientific:2019vic} and the} $1$-$\sigma$ power-law integrated (PI) curve for the LIGO design sensitivity with two years of integrated time in dashed black line for the frequency dependence $f^{2/3}$ \citep{TheLIGOScientific:2016wyq}.}
\label{sgwballlvc}
\end{figure*}

The corresponding estimate of the SGWB power spectrum for the gravitational wave sources in the mass range $[5\,M_\odot, 50\,M_\odot]$, assuming the mass distribution function for both the black hole as $p(\theta)= 1/m^{2.35}$ and merger rates $R(z)$ (given in Eq. \ref{rates}) is shown in Fig. \ref{sgwball} for the frequency range which is relevant for the LVC detectors. This plot indicates that the amplitude of the SGWB signal increases with the increase in the amplitude of the merger rates, and the shape of the signal depends on the redshift distribution of merger rates. For the merger rates which have a lower  contribution from the low redshift Universe ($z<1$), there is a decrease in the amplitude of the SGWB power spectrum at high frequency. This is because the window function $\mathcal{W}(z,f)$ for the high-frequency signal gets more relative contribution from the low redshift Universe (as shown in Fig. \ref{window}) than for the low-frequency SGWB signal. This indicates that SGWB power spectrum is a direct probe of the merger rates of the gravitational wave sources (as also pointed out previously \citep{10.1093/mnras/stz3226, Boco:2019teq, Callister:2020arv}) and can be used to learn about the high redshift astrophysical systems. We also plot the power-law integrated (PI) detector noise for the two years of design sensitivity of LVC by the black dashed line \citep{PhysRevD.88.124032, TheLIGOScientific:2016wyq}. In Fig. \ref{sgwballlvc}, we plot the power spectrum of SGWB signal $\Omega_{gw}(f)$ for the mass range $[5 \, M_{\odot}, 95\,M_{\odot}]$, and consider the distribution function of the gravitational wave source mass as $p(\theta)= 1/m_i^{2.35}$ (for the heavier mass) and $1/m_i$ for the lighter mass. This plot exhibits the variation in the amplitude of the SGWB signal for the different mass distribution function.  {In Fig. \ref{sgwball} and Fig. \ref{sgwballlvc} we also plot the $95\%$ upper bound on the SGWB power spectrum from the observation run of O1+O2 for the spectrum $\Omega_{GW} (f) \propto f^{2/3}$ \citep{LIGOScientific:2019vic}. This indicates that the different models of merger rates considered in this analysis are allowed by the current upper bounds on the SGWB power spectrum.}

 \section{Relating strong lensing event rates to the SGWB signal}\label{lensgwb}
Strong lensing of gravitational waves can lead to magnified strain as we have discussed in Sec. \ref{seclen}. The number of such lensed events depend on the redshift distribution of the merger rate $R(z)$ and can vary by several orders of magnitude with the change in the merger rates as we have shown in Fig. \ref{lensedratesdesign}. So, an independent probe to the high redshift merger rates is going to be useful to estimate the expected lensed event rate. The SGWB signal brings an avenue to measure the high redshift merger rates of the gravitational wave sources. 

The probability of strongly lensed events is rare due to the small value of the lensing optical depth $\tau (\mu, z)$ for large magnification factor $\mu>>1$ (as shown in Fig. \ref{Fig:tau}). As a result, only a few events can get strongly lensed and can be detected as a loud individual event, with the matched-filtering SNR $\rho\geq \rho_{th}$. Several sources which are not lensed and below the detection threshold $\rho< \rho_{th}$ contribute to the SGWB power spectrum $\Omega_{GW}(f)$. As a result, there must exist a strong relationship between the strongly lensed event rate and the strength of the SGWB signal. In this section of the paper, we explore what can be learned about the strongly lensed gravitational wave events using the SGWB signal. 
\begin{figure*}
\centering
\includegraphics[trim={0cm 0.cm 1.cm 0.5cm},clip,width=0.8\textwidth]{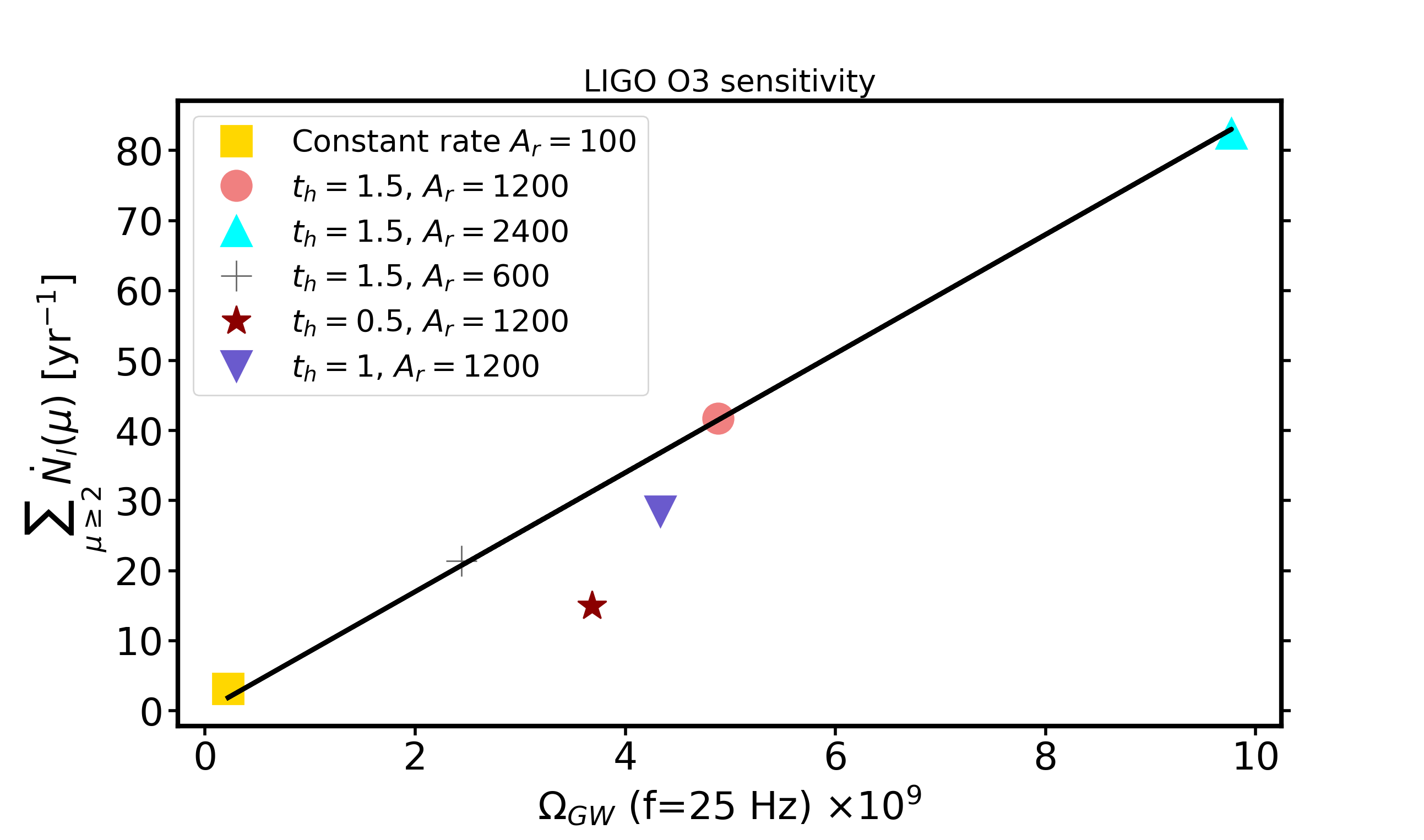}
\caption{The presence of correlations between the SGWB background at $f= 25$ Hz and the total number of expected lensed systems above magnification factor $\mu$ ($\sum_{\mu \geq 2} \dot N_l$) for the LIGO O3 detector sensitivity are shown. The straight line shows the existence of a linear relationship between these two quantities. The departure from the straight line arises due to the combined effect from detector response function $\mathcal{S}(\theta, z, \mu)$ and the merger rate in the low redshift range. Details are explained in the text.}
\label{sgwb-lens-o3}
\end{figure*} 

\begin{figure*}
\centering
\includegraphics[trim={0cm 0.cm 1.cm 0.5cm},clip,width=0.8\textwidth]{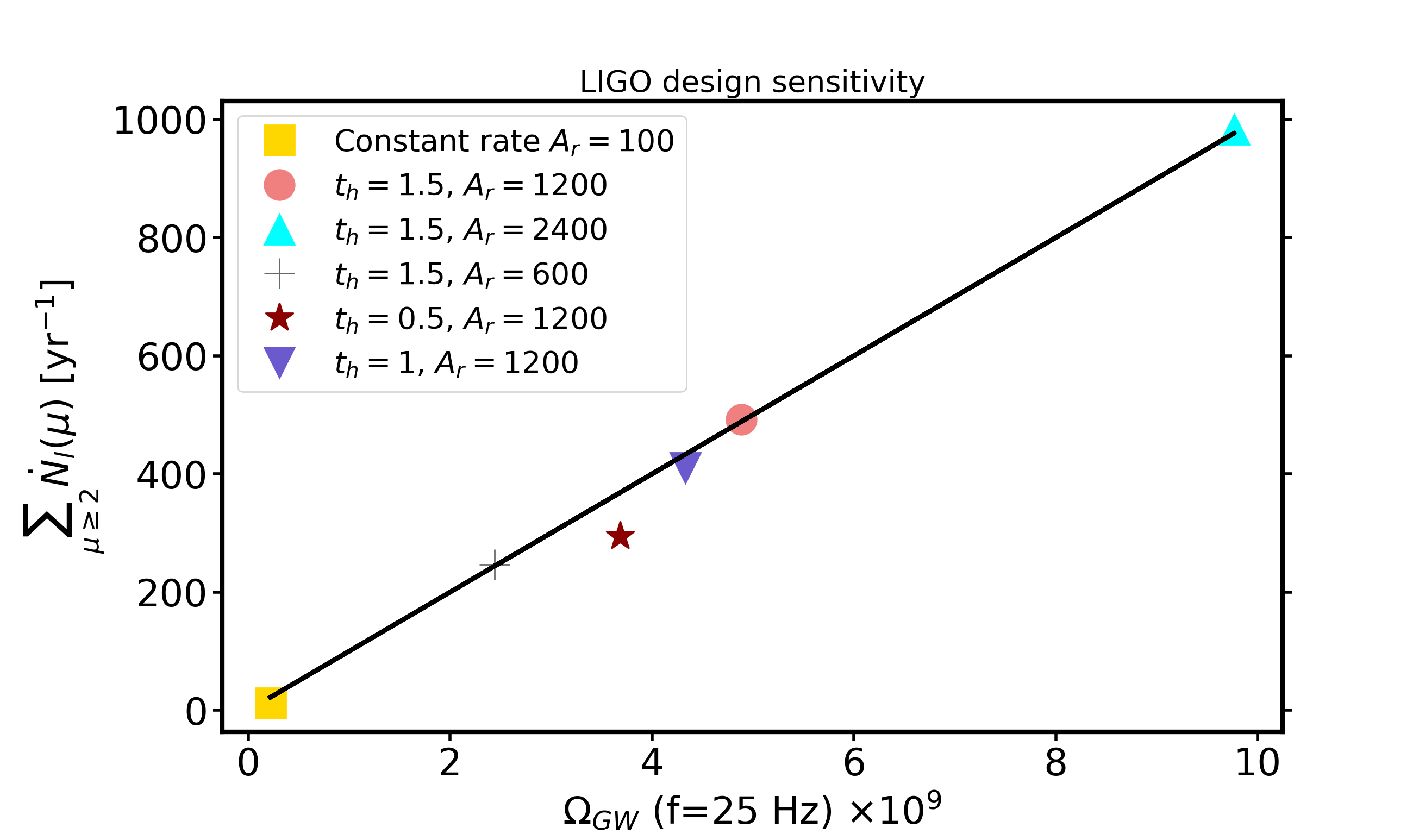}
\caption{The existence of the similar correlation (as shown in Fig. \ref{sgwb-lens-o3}) for the LIGO design sensitivity is shown in this plot. The departure of a few points  from the straight line is less severe than for the case shown in Fig. \ref{sgwb-lens-o3} due to reach of the the detector response function $\mathcal{S}(\theta, z, \mu)$ up to higher redshift than LIGO design sensitivity.}
\label{sgwb-lens}
\end{figure*}  

The comparison of Eq. \eqref{nzl-1} and Eq. \eqref{sgwb-1} shows that both the number of lensed events and the SGWB signal depends on the volume integral of the astrophysical merger rates $R(z)$ and mass distribution of the gravitational wave sources $p(\theta)$. So, using the same merger rate $R(z)$ and mass distribution $p(\theta)$ of the gravitational wave sources, we estimate the total number of detectable lensed events with the advanced LIGO O3 and design sensitivity, and also the expected power spectrum of the SGWB signal. The corresponding plot between total lensed event rate $\sum_\mu \dot N_l (\mu)$ and SGWB power spectrum $\Omega_{GW}(f)$ at frequency $f=25$ Hz is shown in Fig. \ref{sgwb-lens-o3} and Fig. \ref{sgwb-lens} for the O3 sensitivity and design sensitivity respectively \footnote{The choice of $f=25$ Hz is made because detector noise of LVC on the SGWB power spectrum is small at $f=25$ Hz. This is explained in details in Sec. \ref{bounds}.}. This figure shows that the event rate of detectable strongly lensed systems is related to the amplitude of the SGWB power spectrum. This arises from the inevitable correlation of the  gravitational wave mergers from the  common redshift range between SGWB window function $\mathcal{W}(z)$ and the lensing kernel $\mathcal{K}(\theta, \mu, z)\equiv \tau(\mu, z)\mathcal{S}(\theta, \mu, z)$. The deviation around the strong correlation takes place due to the partial overlap of SGWB window function and the lensing kernel for different values of the magnification factor $\mu$ (We explain this in the details in the paragraph after Eq. \ref{thresomg}). The number of detectable lensed systems can be lower than this line due to the detector response function $\mathcal{S}(\theta, \mu, z)$. As a result, this line represents an upper bound on the expected lensed event rates. 
\begin{figure*}
\centering
\includegraphics[trim={0cm 0.cm 0cm 0.cm},clip,width=0.8\textwidth]{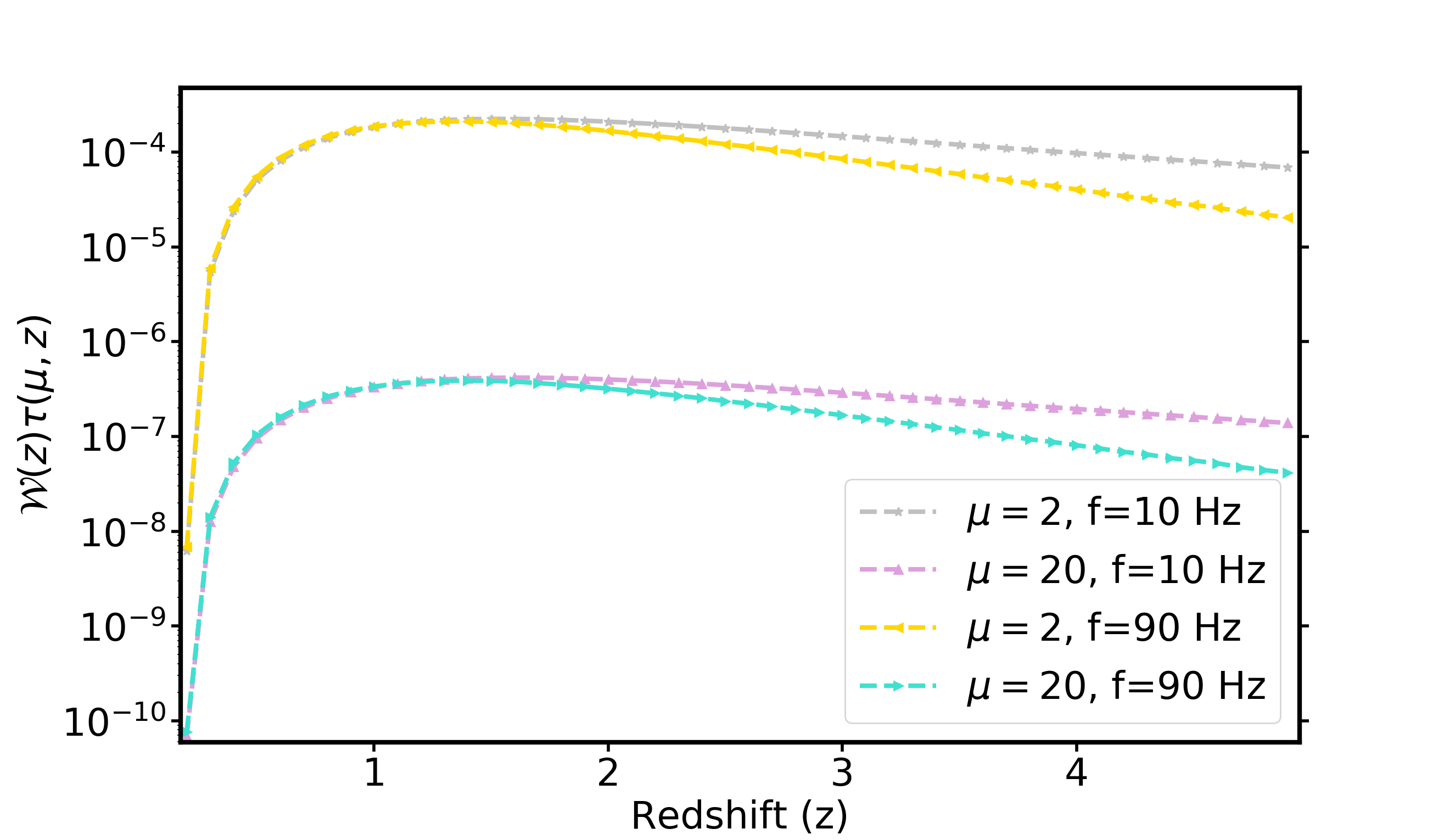}
\caption{The overlap of the SGWB window function at different GW frequencies $\mathcal{W}(z)$ with the lensing optical depth for different magnification factors $\tau(\mu,z)$ is shown as a function of redshift. The SGWB window function for smaller frequencies has a larger contribution from high redshift than higher frequencies, resulting in the deviation shown above for cases with the same magnification factor $\mu$. Variation in $\tau(\mu, z)$ for different values of $\mu$ (see Fig. \ref{Fig:tau}) leads to the variation between $\mu=2$ and $\mu=20$.}
\label{wztau}
\end{figure*} 

The origin of a strong correlation with the total number of lensed event rate $\sum_\mu \dot N_l(\mu, z)$ and the SGWB signal can be understood better, by exploring the overlap between the SGWB window function $\mathcal{W}(z)$ and lensing optical depth $\tau(\mu, z)$ and lensing kernel $\mathcal{K}(\theta, \mu, z)$ for different values of the magnification factor $\mu$. We show the overlap of the lensing optical depth $\tau(\mu, z)$ with the SGWB window function $\mathcal{W}(z)$ in Fig. \ref{wztau} for two gravitational wave frequencies $f=10$ Hz and $f=90$ Hz \footnote{These frequency ranges corresponds to the low SGWB noise values for the LIGO detectors. See the black dashed line in Fig. \ref{sgwball}.} for the magnification factor $\mu=2$, and $\mu=20$. The plot shows that contribution from high redshift is more in the low-frequency range of the SGWB signal, and the contribution from  low redshift is similar for both the frequencies. The contribution is stronger for small values of the magnification factor $\mu$ arising from lensing optical depth $\tau(\mu, z)$. So, the low-frequency SGWB signal can better probe the high redshift merger rates than the high-frequency SGWB signal. 

\begin{figure*}
\centering
\includegraphics[trim={0cm 0.cm 0cm 0.cm},clip,width=.9\textwidth]{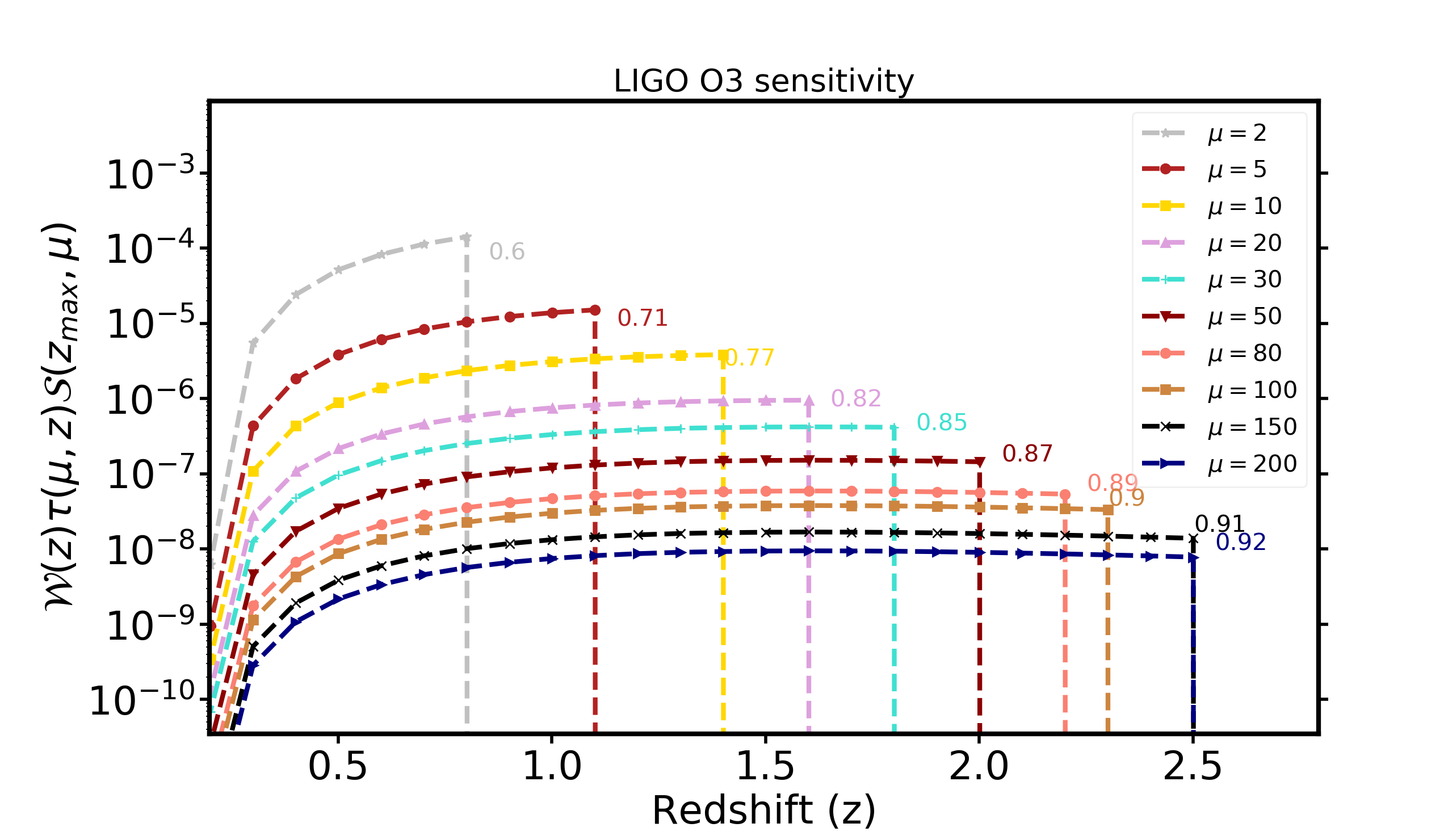}
\caption{We show the overlap in the redshift between the SGWB window function at $f=25$ Hz $\mathcal{W}(z)$, lensing optical depth $\tau(\mu, z)$, and LIGO O3 detector response function for the magnification factor $\mu$ $\mathcal{S}(z_{max, \mu})$ as a function of redshift. This plot shows that for a lower magnification factor, the overlap redshift range is smaller than for the cases with higher magnification factor. This implies the SGWB is more informative about the strongly magnified systems than the weakly magnified systems, which can be observed only up to small values of redshift. The values mentioned alongside every magnification factor $\mu$ denotes the integrated area of the SGWB window function $\mathcal{W}(z)$ up to $z_{max}$. }
\label{wztauszO3}
\end{figure*}

\begin{figure*}
\centering
\includegraphics[trim={0cm 0.cm 0cm 0.cm},clip,width=.9\textwidth]{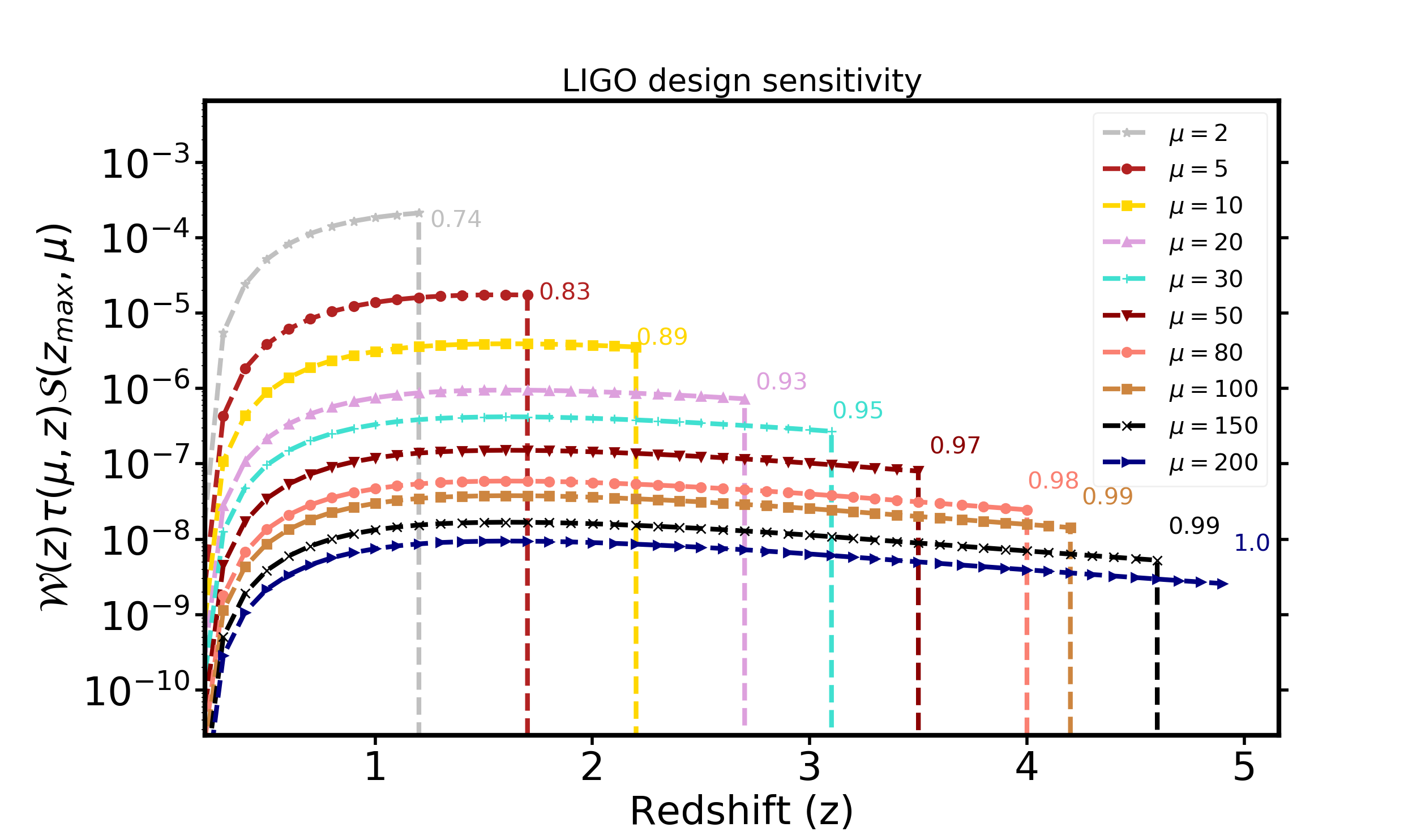}
\caption{Similar to Fig. \ref{wztauszO3}, we obtain the overlapping redshift between the SGWB window function at $f=25$ Hz $\mathcal{W}(z)$, lensing optical depth $\tau(\mu, z)$, and LIGO design detector response function for the magnification factor $\mu$ $\mathcal{S}(z_{max, \mu})$ as a function of redshift. Due to reach of the detector response function $\mathcal{S}(z_{max, \mu})$ up to a higher redshift than O3, the integrated area f the SGWB window function $\mathcal{W}(z)$ is more than  the results obtained for O3.}
\label{wztausz}
\end{figure*}

The overlapping redshift range between the detected lensed events and the SGWB signal at $f=25$ Hz for different magnification factors is shown in Fig. \ref{wztauszO3} and Fig. \ref{wztausz} respectively for the LIGO O3 sensitivity and LIGO design sensitivity. The objects which are highly magnified can be observed up to a higher redshift than the less magnified objects. As a result, the product of the lensing kernel $\mathcal{K}(\theta, \mu, z)$ and the SGWB window function $\mathcal{W}(z)$ goes to zero for values of redshift $z>z_{th}$\footnote{$z_{th}$ is defined as the redshift up to which we can make detection of the gravitational wave signal with an SNR $\rho =8$ (given in Eq. \ref{snrgw}.)}, and the value of $z_{th}$ increases with the increase in the magnification factor. The fraction of the SGWB window function $\mathcal{W}(z)$ which contributes up to redshift $z_{th}$ ($\int_0^{z_{th}} dz \mathcal{W}(z)$) is mentioned, alongside the corresponding cases in Fig. \ref{wztauszO3} and Fig. \ref{wztausz} for LIGO O3 sensitivity, and LIGO design sensitivity respectively. The corresponding part of the SGWB signal, which is going to have an overlap with the lensing kernel, can be written as
\begin{equation}\label{thresomg}
    \Omega^{z_{\mu}}_{GW} (f)= \frac{f}{\rho_c c^2}\int_0^{z_{th}(\mu)} dz \mathcal{W}(f, z) R(z).
\end{equation}
This shows that for the small values of the magnification factor, the fraction of the SGWB window function which can have an overlap with the lensing kernel, is small, resulting in only a partial correlation. On the other hand, for the highly magnified systems, the overlap with the lensing kernel is almost $~100\%$, indicating a strong correlation with the number of lensed event rate. In other words, the deviations from the strong correlation between SGWB amplitude with the total number of lensed events are going to arise from less magnified systems, and hence from the merger rate associated with the low redshift Universe $z<1$. In Fig. \ref{sgwb-lens-o3} and Fig. \ref{sgwb-lens} the deviations from the strong correlation between the lensed event rates and SGWB amplitude happens for those merger rates $R(z)$ (shown in Fig. \ref{rates}) which has a rapid decrease in the number of events in the low redshift $z<1$ are smaller than the case with merger rate constant $A_r=10^2$ Gpc$^{3}$ yr$^{-1}$ (This happens for the cases with $t_h< 1.5$ Gyr).  

\begin{figure*}
\centering
\includegraphics[trim={0cm 0.1cm .3cm 0.5cm},clip,width=0.8\textwidth]{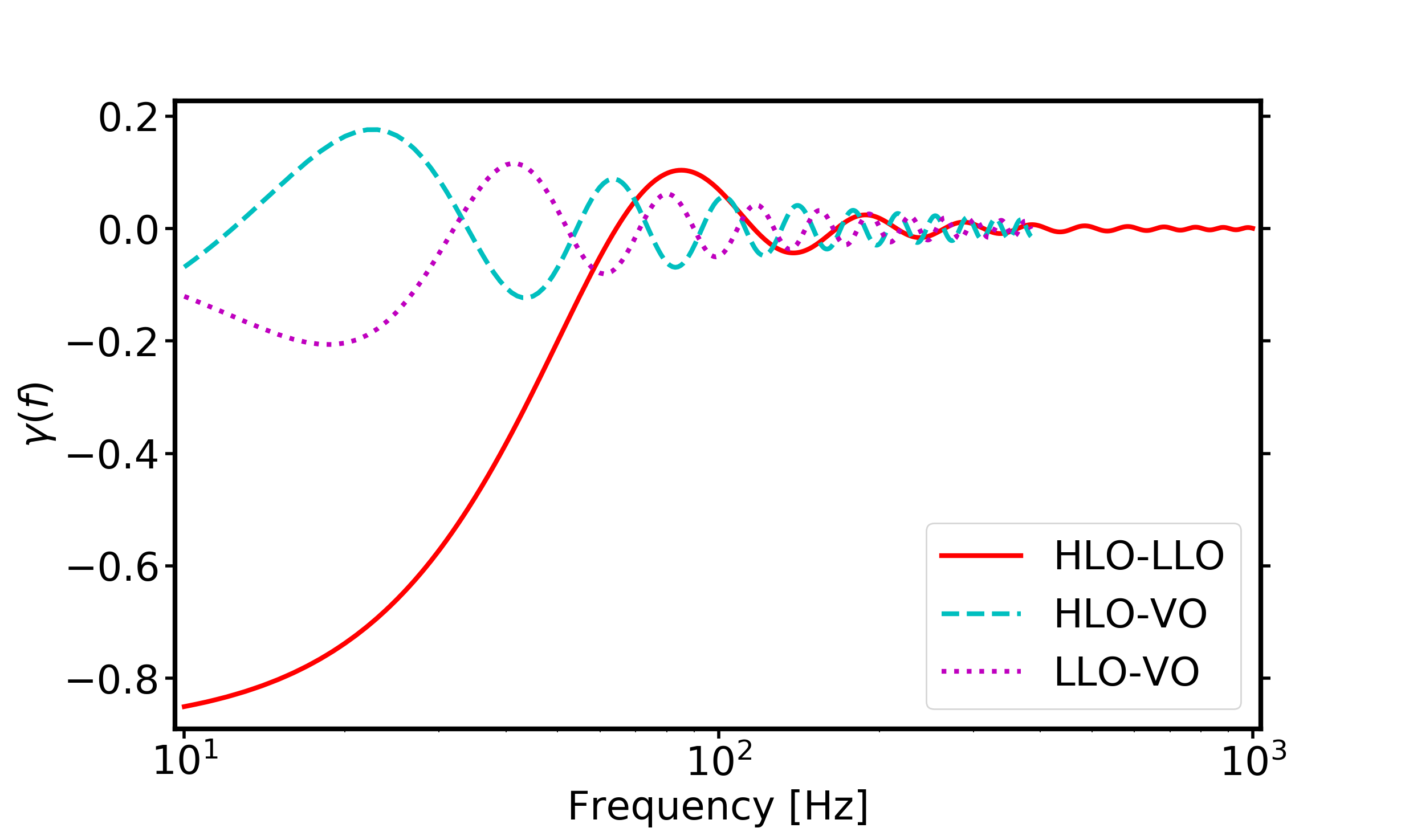}
\caption{We show the normalised overlap reduction function for the Hanford LIGO Observatory (HLO), Livingstone LIGO Observatory (LLO), and Virgo Observatory (VO). The first zero crossing of the overlap reduction function is denoted by $f_{char}$ above which the signal to noise ratio reduces.}
\label{overlapf}
\end{figure*}

The existence of the strong correlation between the lensed event rate and the SGWB power spectrum is going to be useful for predicting  the expected lensed event rate for different magnification factor. By using the measurement from the SGWB power spectrum, we can write the probability distribution of the expected event rate as  
\begin{align}\label{sgwb-len-1}
    \begin{split}
        \mathcal{P}(\dot{N}_l (\mu)| \hat \Omega_{gw}) &=  \int d\theta\, \mathcal{P} (\dot{N}_l (\mu)| R(z), \theta)\mathcal{P}(R(z)|\hat \Omega_{gw}, \theta) \Pi(\theta) ,
            \end{split}
\end{align}
where, $\Pi(\theta)$ is the prior on the gravitational wave source parameters $\{\theta\}$,  the posterior on the event rates $\mathcal{P}(R(z)| \hat \Omega_{gw})$ can be obtained using the measurement of the SGWB power spectrum $\hat \Omega_{gw}$ 
 \begin{align}\label{sgwb-len-2}
    \begin{split}
       \mathcal{P}(R(z)| \hat \Omega_{gw}, \theta) &=  \mathcal{L} (\hat \Omega_{gw}| R(z), \theta)  \Pi(R(z)|\theta),
                    \end{split}
\end{align}
where,  $\Pi(R(z)|\theta)$ is the prior on the merger rate of the gravitational wave sources given the GW source parameters  {and $\mathcal{L} (\hat \Omega_{gw}| R(z), \theta)\propto \exp{(-\chi^2/2)}$ is the likelihood with}   
 \begin{align}\label{sgwb-len-3}
    \begin{split}
    \chi^2=& \int_{0}^{T_{obs}} dt \int_{-\infty}^{\infty} df  \bigg(\frac{3H_0^2}{10\pi^2}\bigg)^2\\ & \times \sum_{IJ}\frac{\gamma^2_{IJ}(\hat \Omega_{IJ}(f) - \Omega_{gw} (f, R(z), \theta))^2}{f^6S_{n_I}(f)S_{n_J}(f)},
                           \end{split}
\end{align}
where integration is performed over the observation time $T_{obs}$, and the frequency of the gravitational wave signal, the summation over the indices (I, J) denotes the summation over all the combinations of gravitational wave detector pairs,  $\Omega_{gw} (f, R(z), \theta)$ is the theoretical power spectrum of the SGWB, $S_{n_I}(f)$ is the detector noise power spectrum for  detector $I$, and $\gamma_{IJ}$ is the normalised overlap reduction function shown in Fig. \ref{overlapf} for all the combinations of the LIGO and Virgo detectors \footnote{The overlap reduction function can be downloaded from this link  \href{https://dcc.ligo.org/public/0022/P1000128/026/figure1.dat}{P1000128/026}.}.  {The inclusion of the prior on the gravitational wave source parameters is crucial as the shape and amplitude of the SGWB power spectrum varies for the different mass distributions (as shown in Fig. \ref{sgwball} and Fig. \ref{sgwballlvc}) and can also show mild variation for including the contribution from spin \citep{Zhu:2011bd}.} 

By including the measurements of the SGWB data, we can obtain the posterior on the merger rates of the compact objects, and use it as a prior for estimating the rate of the number of lensed events above a magnification factor $\mu$.  Detection of the SGWB signal is going to provide a bound on the number of expected lensed events according to Eq. \eqref{sgwb-len-1}. Even in the absence of a detection of the SGWB signal within the next few years, the existence of an upper bound in the SGWB will lead to an upper bound on the total number of expected lensed events. This will help in reducing the uncertainty associated with the lensed event rate due to the uncertainty in the high redshift merger rate. In this analysis, we have discussed only the lensing event rates and the corresponding SGWB signal for BBHs. But gravitational wave sources such as binary neutron stars (BNSs) and neutron star black hole (NS-BH) systems can also contribute to the lensed events and in the SGWB signal. For both of these kinds of systems, sources can be detected only up to low redshift than BBHs and can rarely exhibit high magnification. The corresponding SGWB signal in the low-frequency range ($< 100$ Hz) is also going to smaller than the contribution from BBHs \citep{LIGOScientific:2018mvr}. We do not discuss these sources due to the large uncertainty in the event rate. But they can be easily included for the joint estimation of lensing event rate and SGWB signal proposed in this paper.

\section{Possible bounds on the lensed event rates from SGWB measurements}\label{bounds}

The SGWB signal is a diffuse background of spatially unresolved sources, out of which a few events can get strongly lensed as discussed in the previous section. If the SGWB signal can be measured with a high spatial and temporal resolution, then one can relate the SGWB signal strength with the number of lensed events coincident in both space and time. However for the weak SGWB signal, spatially and temporally-averaged SGWB signal is going to be more informative about the global event rate which can be related to the lensing event rate (as discussed in Sec. \ref{lensgwb}). In the remaining of this section, we will only discuss  the all-sky averaged SGWB signal.

The SGWB signal can be measured by cross-correlating the short time Fourier transform gravitational wave signal $d_i(f, t)= h_i(f,t) + n_i(f,t)$ \footnote{The short-time Fourier transform of the sky signal is defined as $d(t, f)=\int_{t- \tau/2}^{t+\tau/2} d(t') e^{-2\pi ift'} dt'$.} between two different pair of detectors (I,J) as \footnote{Here the $\langle.\rangle$ is performed over time.} \citep{Allen:1997ad,Mitra:2007mc, Thrane:2009fp, Talukder:2010yd,Romano:2016dpx}
 \begin{align}\label{sgwb-len-3}
    \begin{split}
     \Omega^d_{IJ}(f) \equiv & \langle{d_I(f,t)d^*_J(f',t)}\rangle =\delta (f-f') \langle h_I(f,t)h^*_J(f', t) \rangle \, \\& + \cancel{\langle n_I(f,t)h^*_J(f', t) \rangle}\,  +   \cancel{\langle h_I(f,t)n^*_J(f', t) \rangle}   \\& +\, \cancel{\langle n_I(f,t)n^*_J(f', t) \rangle},
                           \end{split}
\end{align}
where all the terms which include the detector noise get cancelled out as these are not correlated with the signal and neither with the noise of the other detector. The observed gravitational wave strain $h_I (f, t)$ is a convolution of the detector response function $F^p_I(\hat n, t)$ and the true sky signal $h_p^s$, and can be expressed as \citet{PhysRevD.46.5250, PhysRevD.48.2389, Allen:1997ad}
 \begin{eqnarray}\label{es1a}
h_I(t, \hat n) = \int_{-\infty}^{\infty} df\int d^2\hat n\, F^p_I(\hat n, t)e^{2\pi i f(t- \vec x_I. \hat n/c)}h^s_{p} (f, \hat n).
\end{eqnarray}
\begin{figure*}
\centering
\includegraphics[trim={0cm 0.cm 1.cm 0.5cm},clip,width=0.8\textwidth]{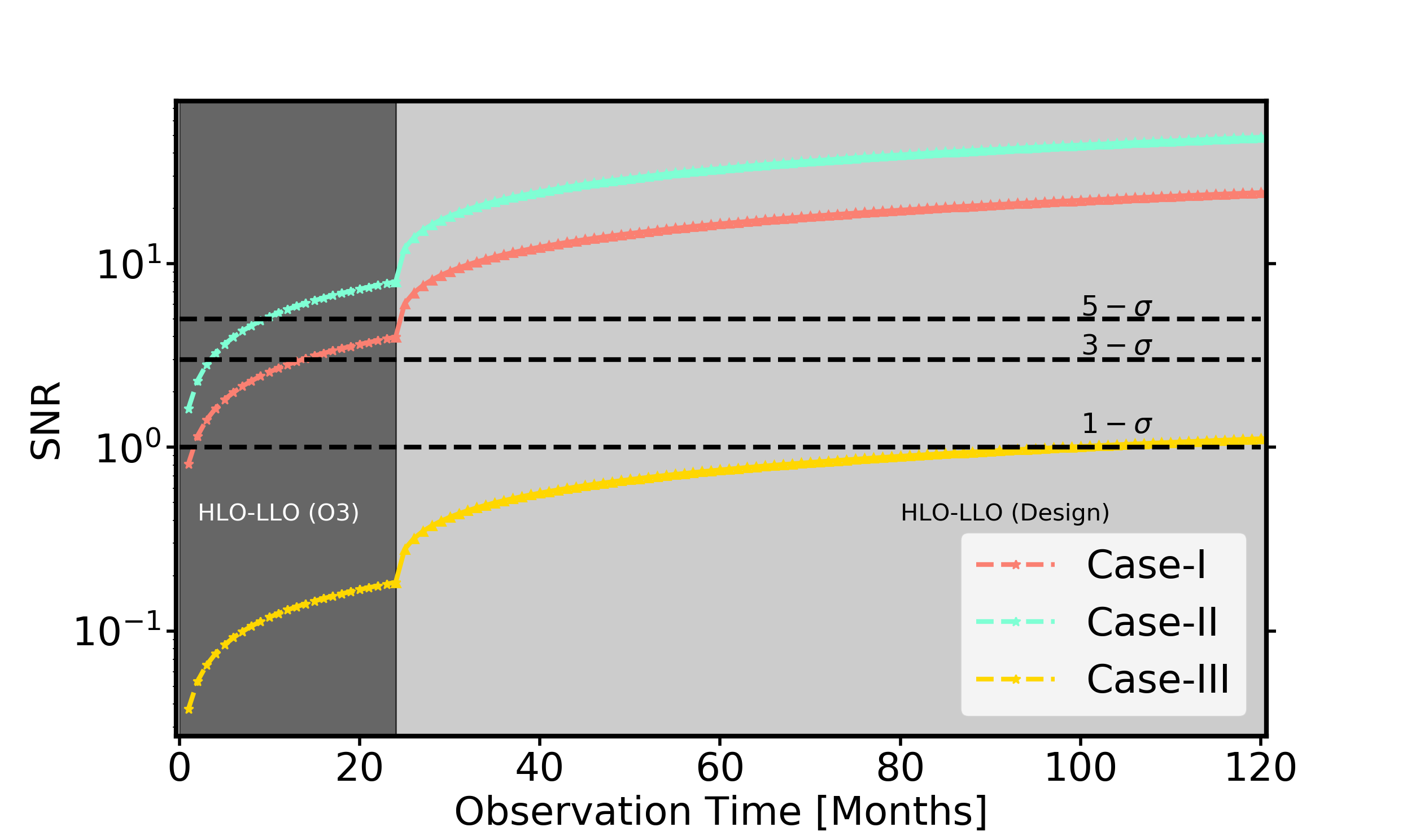}
\caption{We show the possible signal-to-noise ratio (SNR) on the SGWB signal for the detector network Hanford LIGO observatory (HLO) and Livingston LIGO Observatory (LLO) for the O3 detector sensitivity (in dark grey), and the future LIGO design sensitivity (in light grey). Case-I, II and III denotes three different event rates $A_r= 1200$ Gpc$^{-3}$ yr$^{-1}$ and $t_h= 1.5$ Gyr, $A_r= 2400$ Gpc$^{-3}$ yr$^{-1}$ and $t_h= 1.5$ Gyr, and constant rate with $A_r= 100$ Gpc$^{-3}$ yr$^{-1}$ respectively with probability distribution of the mass function $p(\theta)=1/m^{2.35}$ for both the masses in the range $[5\,M_\odot, 50\, M_\odot]$ corresponding to the signal shown in Fig. \ref{sgwball}. The black dashed lines denotes the $1-\sigma$, $3-\sigma$, and $5-\sigma$ error bars on the SGWB signal.}
\label{snrf}
\end{figure*}
So, the observed cross-correlation signal $\Omega^d_{IJ}(f)$ is related to the true SGWB signal by the relation 
 \begin{eqnarray}\label{es1b}
\Omega^d_{IJ}(f,t)= \int d^2 \hat n \gamma(\hat n, f, t)\Omega^s(\hat n, f, t),
\end{eqnarray}
where  $\gamma(\hat n, f, t)$ is the overlap reduction function \citep{PhysRevD.48.2389, PhysRevD.46.5250}
 \begin{eqnarray}\label{es3a}
\gamma(\hat n, f, t)= \frac{1}{2}F^p_i (\hat n, t)F^p_j (\hat n, t)e^{2\pi i f\hat n.(\vec x_i- \vec x_j)/c}.
\end{eqnarray}
The normalized overlap reduction function $\gamma(\hat n, f, t)$ for Hanford LIGO Observatory (HLO) and Livingston LIGO Observatory (LLO), and Virgo are shown in Fig. \ref{overlapf}\footnote{The overlap reduction is given in the following link \href{https://dcc.ligo.org/public/0022/P1000128/026/figure1.dat}{P1000128/026}}. The first zero crossing of the overlap reduction function $\gamma(f)$ takes place at $f=f_{char} \equiv c/2D$ which depends on the distance between the pair of detectors $D$ \citep{PhysRevD.48.2389, PhysRevD.46.5250}. For the frequency range $f>f_{char}$ overlap reduction function oscillates around zero, which diminishes the observed cross-correlation signal given in Eq. \eqref{es1b}. The spatial fluctuations in the SGWB will also exhibit spatial fluctuations as expressed in Eq. \ref{es1b} and can be estimated using the methods developed by \citep{Mitra:2007mc, Thrane:2009fp}. The angular resolution of SGWB signal is diffraction limited $\Delta \theta= c/2fD = f_{char}/f$. However, for $f>f_{char}$ the observed signal $\Omega^d_{IJ}(f,t)$ is close to zero. As a result, only the signal in $\Delta \theta \gtrsim 1$ are measurable with high signal to noise ratio (SNR) \citep{10.1093/mnras/stz3226}.

The measurability of the SGWB signal depends on the power spectrum of the pair of detectors $S_{n_I}(f)S_{n_J}(f)$ for the pair of detector $I$, and $J$. Using the LIGO O3 and LIGO designed sensitivity of HLO, and LLO detector pair, we obtain the signal to noise ratio (SNR)  using the relation
\begin{align}\label{snr-1}
    \begin{split}
          SNR= \bigg[\int_0^T dt & \sum_{I, J>I}^N\bigg(\frac{3H_0^2}{10\pi^2}\bigg)^2  \int_{-\infty}^\infty df \frac{\gamma_{IJ}^2(f)\Omega^2_{GW}(f)}{f^6 S_{n_I}(f)S_{n_J}(f)}\bigg]^{1/2}, 
     \end{split}
 \end{align}
where the summation is made over all available detector combinations and $T$ denotes the observation time. 
  \begin{figure*}
\centering
\includegraphics[trim={0.cm 0.cm 1.cm 0.5cm},clip,width=.8\textwidth]{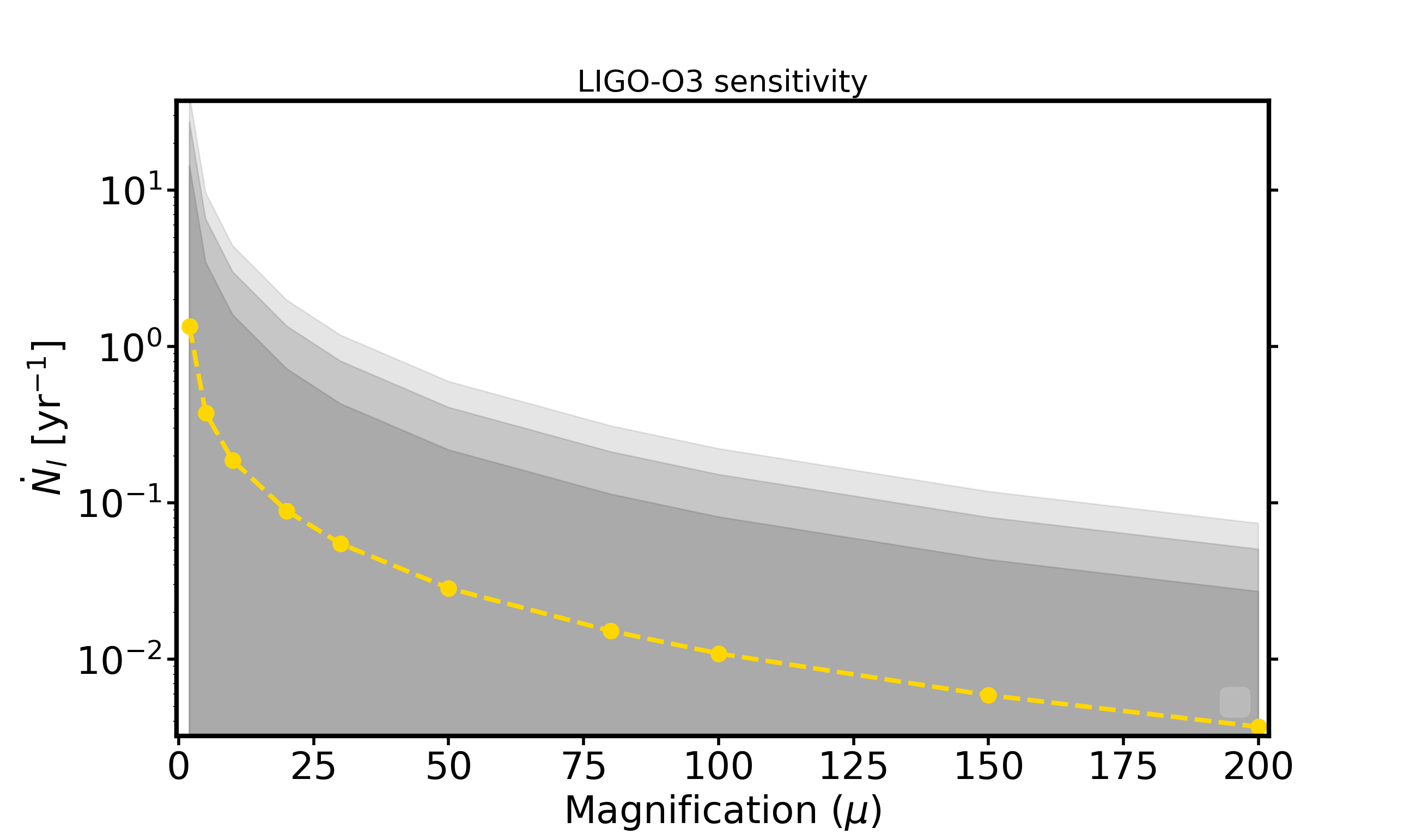}
\caption{Possible $1-\sigma$ (dark grey) to $3-\sigma$ (light grey) bounds on the rate of strongly lensed events as a function of the magnification factor $\mu$ are shown for LIGO O3 design sensitivity. The yellow line denotes the lensing event rate for the benchmark model (case of constant rate $A_r=10^2$ Gpc$^{-3}$ yr$^{-1}$) is also shown in Fig. \ref{lensedratesO3} and Fig. \ref{lensedratesdesign}.} 
\label{nlf}
\end{figure*}

\begin{figure*}
\centering
\includegraphics[trim={0.cm 0.cm 1.cm 0.5cm},clip,width=.8\textwidth]{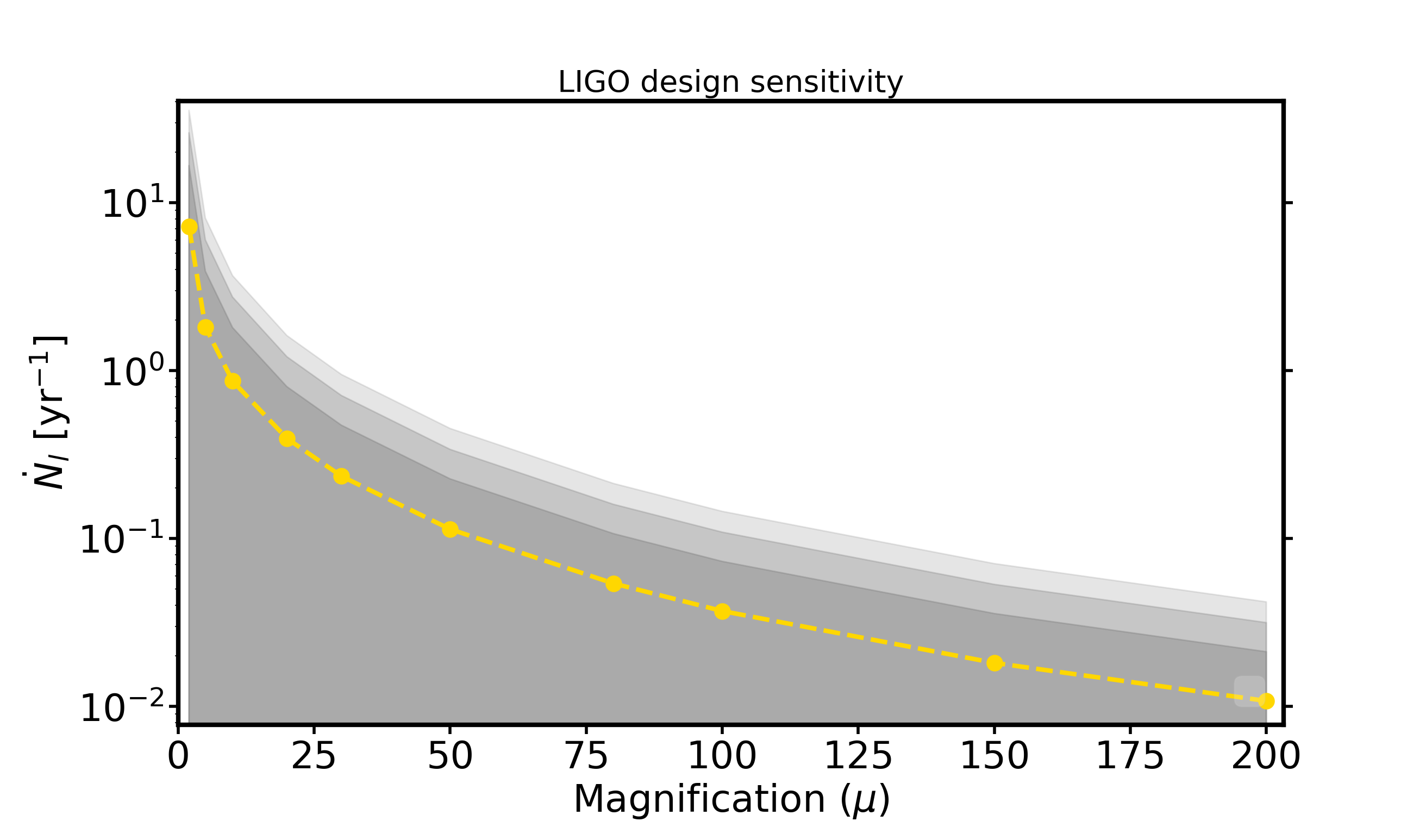}
\caption{Bounds on the lensed event rates (similar to Fig. \ref{nlf}) for the LIGO design sensitivity are shown as a function of the magnification factor $\mu$. The yellow line denotes the lensing event rate for the benchmark model (case of constant rate $A_r=10^2$ Gpc$^{-3}$ yr$^{-1}$) is also shown in Fig. \ref{lensedratesO3} and Fig. \ref{lensedratesdesign}.}
\label{nlfd}
\end{figure*}

The SGWB has not yet been detected from the O1 and O2 observations by the LVC \citep{LIGOScientific:2019vic}. The currently existing $95\%$ bounds on the SGWB signal for the frequency dependence $f^{2/3}$ from the combined O1 and O2 analysis are $4.8\times 10^{-8}$ and $3.0\times 10^{-8}$ for the uniform prior and log-normal prior respectively \citep{LIGOScientific:2019vic}. These bounds on the SGWB are weak and as a result, even very large event rates are still allowed, including all the merger rates considered in this analysis. So, we do not explore the impact of the O1 and O2 runs on the lensing event rates. For the currently ongoing LVC run with O3 sensitivity, and for the future LIGO designed sensitivity\footnote{The noise curves can be downloaded from the following link \href{https://dcc-lho.ligo.org/LIGO-T2000012/public}{https://dcc-lho.ligo.org/LIGO-T2000012/public}}, we obtain the detectability for the different SGWB signal corresponding to different merger rates. We plot the SNR as a function of the observation time in Fig. \ref{snrf} for three different merger rates, (i) the highest merger rate (case-I: $A_r=2400$ Gpc$^{-3}$ yr$^{-1}$, $t_h= 1.5$ Gyr), (ii) merger rate following the previous analysis \citep{Broadhurst:2018saj, Broadhurst:2019ijv,Broadhurst:2020moy} (case-II: $A_r=1200$ Gpc$^{-3}$ yr$^{-1}$, $t_h= 1.5$ Gyr), and the fiducial case currently agrees with O1 and O2 gravitational wave data  \citep{LIGOScientific:2018mvr} (case-III: constant with $A_r=100$ Gpc$^{-3}$ yr$^{-1}$)
The region shaded in dark grey indicates the SNR possible with O3 HLO-LLO, and the region shaded in light grey indicates the SNR possible from the designed sensitivity of HLO-LLO. This plot indicates that cases with high merger rates case-I and case-II can be detectable with the currently ongoing O3 detector sensitivity, as a result, measurement of such a signal can place a direct probe to the expected lensed event rate as shown in Fig. \ref{lensedratesO3} at the $\approx 8 \sigma$ and $\approx 4 \sigma$ respectively. In other words, non-detection of the SGWB signal from the O3 runs will be able to constrain these models (case-I and case-II). For case-III, a one sigma bound is possible with the design sensitivity, which makes it possible to bound the high redshift merger rates of gravitational wave sources.

For the fiducial model of the merger rate (case-III) \citep{LIGOScientific:2018mvr}, we obtain the bounds on the merger rate for the LIGO O3 and LIGO design sensitivity in Fig. \ref{nlf} and Fig. \ref{nlfd} respectively. In both these figures, we plot the expected $1-\sigma$, $2-\sigma$, and $3-\sigma$ upper bound on the lensed event rates in different shades of grey colour. These plots indicate that with the combination of the SGWB data, we will be able to significantly constrain the highly magnified lensed events, even in the absence of a detection of the SGWB signal. Detection of SGWB will be a direct probe to the merger rate, and hence limits on the lensed event rates can be inferred. In the absence of any detection of the SGWB signal from LIGO O3, the bound on strongly lensed event rate for $\mu>200$ is about $0.1$ per year. This can be improved to $\sim 0.05$ per year with the LIGO design sensitivity. In this analysis, we have not considered any SGWB signal arising from  cosmological sources such as from inflation \citep{Starobinsky:1979ty,Turner:1996ck, Martin:2013nzq}, cosmic strings  \citep{Kibble:1976sj,Damour:2004kw, Ringeval:2017eww}, phase transitions \citep{Kosowsky:1992rz,Kamionkowski:1993fg}, standard model \citep{Watanabe:2006qe}, etc. However, these signals are expected to be weaker than the SGWB signal from the astrophysical background for the standard scenarios. So, we do not consider the contributions from the cosmological SGWB signal.

\section{Conclusions}\label{conc}

The predicted level of strong lensing of gravitational wave sources depends on several uncertain factors, including the level of magnification, the source redshift distribution and most importantly, the astrophysical merger rate over cosmic history. As the lensing optical depth is small for high magnification factors, a large event rate is required in the high redshift universe to produce a few lensed events every year, as we have shown in Fig. \ref{lensedratesO3} and Fig. \ref{lensedratesdesign} for LIGO O3 sensitivity and LIGO design sensitivity respectively.  

We have developed a new avenue to limit the uncertain event rate in the high redshift universe  using the SGWB signal. We have found that if a significant number of lensed events is currently present among the current LVC detections, then this must lead to a relatively high level of the SGWB background from the integrated universal population of binary mergers.  We predict a tight correlation between the number of strongly lensed events and the expected level of the SGWB from the common astrophysical origins of both these classes of GW signals, where only a small fraction of the SGWB is from currently detectable individually loud lensed sources. Existence of a strong correlation between both these signals, (as shown in Fig. \ref{sgwb-lens-o3} and Fig. \ref{sgwb-lens}) can lead to inferring the expected signal of one from the other. However, we can expect variation around the strong correlation due to partial overlap of the SGWB window function with the strong lensing optical depth and detector response function (defined in Sec. \ref{seclen}). The partial degeneracy happens mainly for lower redshift mergers, resulting in a higher relative uncertainty in the allowed numbers of strong lensing events of relatively low magnification. The highly magnified events arising from high redshift have maximal overlap with the SGWB window function. 

With the currently ongoing LIGO O3 detector sensitivity, and for the future LIGO design sensitivity, we provide the expected bounds on the lensed event rates for the case of constant binary merger rate consistent with the LVC observations \citep{LIGOScientific:2018mvr}. Even in the absence of detection of the  SGWB signal, we can obtain an upper bound on the number of lensed events as shown in Fig. \ref{nlf} and Fig. \ref{nlfd}. In the vents of an actual detection of the SGWB background level then we have shown that we can infer a rate for lensed events due to the strong correlation between both signals. We can also say that for the relatively high merger rate required to explain most of the LVC binary detections as lensed events \citep{Broadhurst:2018saj, Broadhurst:2019ijv, Broadhurst:2020moy} that there is the prospect soon of detection of the SGWB from the currently ongoing O3 observations, at a level of  $\geq 3\sigma$ from the ongoing LIGO O3 sensitivity. In any case, the stricter upper limits now feasible for the SGWB have the prospect of placing usefully interesting limits on the nature of cosmological gravitational wave sources at substantial redshifts, that will limit the maximum allowable strong lensing event rate, independent of the currently ambiguous inferable rates that sensitively depend on the nature of the binary source population, in particular the unknown form of the high mass distribution of GW events.

This new approach proposed in this work should be useful for understanding the lensed event rates both for the current generation of gravitational wave detectors, and also for the next generation of gravitational wave detectors such as LISA \citep{2017arXiv170200786A}, Cosmic Explorer \citep{Reitze:2019iox}, and the Einstein Telescope \citep{Punturo:2010zz} that will explore a much wider range of GW frequencies. By combining the measurements of the SGWB signal, a better understanding of the expected number of strongly lensed event rate will be possible. 

\section*{Availability of data}
The data underlying this article will be shared on reasonable request to the corresponding author. 

\section*{Acknowledgement}
 S.M. would like to thank Samaya M. Nissanke, and Benjamin D. Wandelt for useful discussions on this work. Authors would like to thank Ajit Kumar Mehta and King Chun Wong for comments on the draft. The results of this analysis are carried out at the Horizon cluster hosted by Institut d'Astrophysique de Paris. We thank Stephane Rouberol for smoothly running the Horizon cluster. The work of SM is supported by the Labex ILP (reference ANR-10-LABX-63) part of the Idex SUPER,  received financial state aid managed by the Agence Nationale de la Recherche, as part of the programme Investissements d'avenir under the reference ANR-11-IDEX-0004-02, and also by the research program Innovational Research Incentives Scheme (Vernieuwingsimpuls), which is financed by the Nether- lands Organization for Scientific Research through the NWO VIDI Grant No. 639.042.612-Nissanke. JMD acknowledges the support of project PGC2018-101814-B-100 (MCIU/AEI/MINECO/FEDER, UE) Ministerio de Ciencia, Investigaci\'on y Universidades.  This project was funded by the Agencia Estatal de Investigaci\'on, Unidad de Excelencia Mar\'ia de Maeztu, ref. MDM-2017-0765. In this analysis, we have used the  following packages: IPython \citep{PER-GRA:2007}, Matplotlib \citep{Hunter:2007},  NumPy \citep{2011CSE....13b..22V}, and SciPy \citep{scipy}. The authors would like to thank the LIGO-Virgo scientific collaboration for providing the noise curves, and overlap reduction function for the network of detectors. LIGO is funded by the U.S. National Science Foundation. Virgo is funded by the French Centre National de Recherche Scientifique (CNRS), the Italian Istituto Nazionale della Fisica Nucleare (INFN) and the Dutch Nikhef, with contributions by Polish and Hungarian institutes.

\bibliographystyle{mnras}
\bibliography{main}
\label{lastpage}

\end{document}